\pdfoutput=1
\documentclass[showpacs,preprintnumbers,10pt,twocolumn]{revtex4}%
\usepackage{amssymb}
\usepackage{amsfonts}
\usepackage{amsmath}
\usepackage{graphicx}
\usepackage{times}
\usepackage{dcolumn}
\usepackage{bm}
\usepackage{revsymb}%
\begin{document}
\title{Impulse-induced localized control of chaos in starlike networks}
\author{Ricardo Chac\'{o}n $^{1}$, Faustino Palmero $^{2}$ and Jes\'{u}s Cuevas-Maraver$^{3}$}
\affiliation{$^{1}$Departamento de F\'{\i}sica Aplicada, E.I.I., Universidad de
Extremadura, Apartado Postal 382, E-06006 Badajoz, Spain and Instituto de
Computaci\'{o}n Cient\'{\i}fica Avanzada (ICCAEx), Universidad de Extremadura,
E-06006 Badajoz, Spain}
\affiliation{$^{2}$Grupo de F\'{\i}sica No Lineal, Departamento de F\'{\i}sica Aplicada I,
Escuela T\'{e}cnica Superior de Ingenier\'{\i}a Inform\'{a}tica, Universidad
de Sevilla, Avda Reina Mercedes s/n, E-41012 Sevilla, Spain}
\affiliation{$^{3}$Grupo de F\'{\i}sica No Lineal, Departamento de F\'{\i}sica Aplicada I,
Escuela Polit\'{e}cnica Superior, Universidad de Sevilla, Virgen de \'{A}frica
7, 41011 Sevilla, Spain and Instituto de Matem\'{a}ticas de la Universidad de Sevilla (IMUS), Edificio
Celestino Mutis, Avda Reina Mercedes s/n, E-41012 Sevilla, Spain}
\date{\today}

\begin{abstract}
Locally decreasing the impulse transmitted by periodic pulses is shown to be a
reliable method of taming chaos in starlike networks of dissipative nonlinear
oscillators, leading to both synchronous periodic states and equilibria
(oscillation death). Specifically, the paradigmatic model of damped kicked
rotators is studied in which it is assumed that when the rotators are driven
synchronously, i.e., all driving pulses transmit the same impulse, the
networks display chaotic dynamics. It is found that the taming effect of
decreasing the impulse transmitted by the pulses acting on particular nodes
strongly depends on their number and degree of connectivity. A theoretical
analysis is given explaining the basic physical mechanism as well as the main
features of the chaos-control scenario.

\end{abstract}

\maketitle

\section{\bigskip INTRODUCTION}

Controlling the dynamical state of a complex network is a fundamental problem
in science [1-5] with many potential applications, including neuronal
disorders in brain networks [6] and evaluation of risks in financial markets
[7]. While most of these works consider networks of linear systems [1,2,5],
only recently has the general and richer case of networks of nonlinear systems
[4] started to be investigated. Here we are interested in controlling networks
in the sense of driving the network from a subset of particular chaotic
initial states to a subset of final (stable) regular states with the view of
not merely identifying driver nodes but also their relative effectiveness as
well as obtaining estimates of the regions in parameter space where suitable
control signals are effective. Regarding the control (suppression and
enhancement) of chaotic states, which is of fundamental interest partly
because of the ubiquity of chaos in nature, including man-made systems, and
partly because of its practical relevance [8-10], it has been shown that the
application of judiciously chosen periodic external excitations is a reliable
procedure for taming chaos in diverse coupled systems such as arrays of
electrochemical oscillators [11], Frenkel-Kontorova chains [12], and recurrent
neural networks [13], just to cite a few instances. Most studies of coupled
nonlinear systems subjected to external excitations have focused on either
global (all-to-all) or local (homogeneous) diffusive-type coupling, while
little attention has been paid to the possible influences of a heterogeneous
connectivity on the regularization of a network's dynamics.

Many diverse real-world networks exhibit heterogeneous connectivity in the
form of a scale-free topology [14] which means that just a small set of nodes
are highly connected $-$the so-called hubs$-$ while the rest of the nodes have
few connections. Since starlike structures are the main motifs of scale-free
networks, here we study the control of chaos in starlike networks of
dissipative driven nonlinear oscillators, expecting that the main features of
such an impulse control scenario may be extensible to the case of scale-free
networks. This is a major motivation for the present work. Chaos control and
synchronization are deeply related phenomena [9], and there have been recent
studies of diverse synchronization phenomena in oscillator networks with
starlike couplings [15-18].

\begin{figure}[tbp]
\begin{tabular}{c}
\includegraphics[width=3cm]{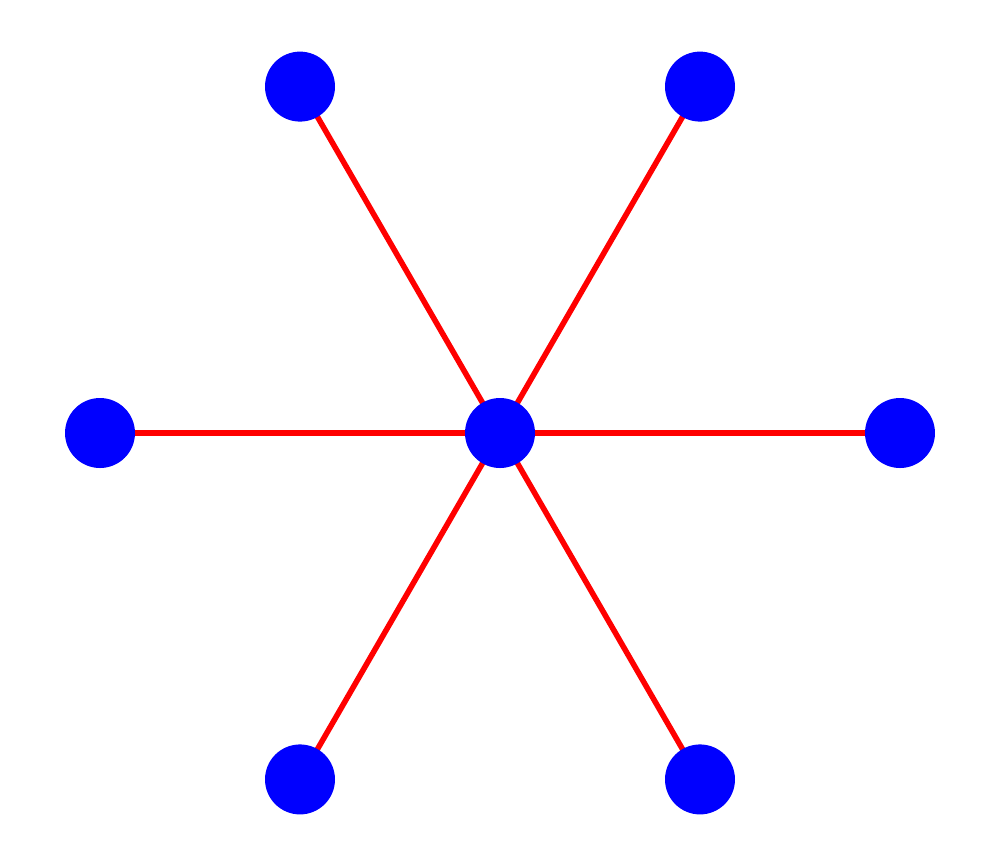} \\
\includegraphics[width=6.5cm]{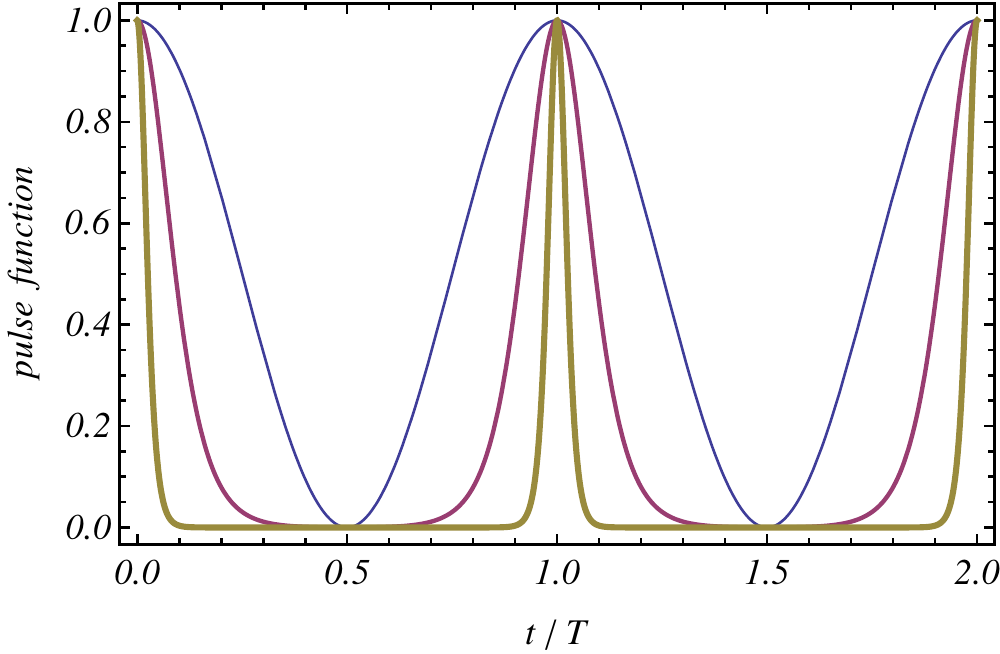} \\
\end{tabular}%
\caption{(Color online) Top: Schematic representation of a starlike network of
$N=7$ rotators. Bottom: Pulse function $p(t;T,m)\equiv\operatorname{cn}%
^{2}\left[  2K(m)t/T;m\right]  $ [cf. Eq.~(1)] versus $t/T$ for $m=0$ (thin
line), $m=0.999$ (medium line), and $m=1-10^{-14}$ (thick line). The
quantities plotted are dimensionless.}
\label{fig1}
\end{figure}

We shall consider a topology-induced chaos-control scenario in starlike
networks of dissipative non-autonomous systems subjected to local
chaos-suppressing (CS) external excitations. Specifically, the findings will
be discussed through the analysis of starlike networks of $N$ damped kicked
rotators (DKRs) -- see Fig.~\ref{fig1} top. This system is sufficiently simple to allow
analytical predictions while retaining the universal characteristics of a
dissipative chaotic system. The complete model system reads%
\begin{align}
\overset{..}{x}_{H}+\operatorname{cn}^{2}\left(  \Omega_{H}t;m_{H}\right)
\sin x_{H}  &  =-\delta\overset{.}{x}_{H}+\lambda\sum_{i=1}^{N-1}\sin\left(
x_{i}-x_{H}\right)  ,\nonumber\\
\overset{..}{x}_{i}+\operatorname{cn}^{2}\left(  \Omega_{i}t;m_{i}\right)
\sin x_{i}  &  =-\delta\overset{.}{x}_{i}+\lambda\sin\left(  x_{H}%
-x_{i}\right)  , \tag{1}%
\end{align}
$i=1,...,N-1$, and where all variables and parameters are dimensionless:
$\Omega_{H,i}=\Omega_{H,i}\left(  T,m_{H,i}\right)  \equiv2K(m_{H,i})/T$, $T$
is the common excitation period, $\delta$ is the damping coefficient,
$\lambda$ is the coupling constant, $\operatorname{cn}\left(  \cdot;m\right)
$ is the Jacobian elliptic function of parameter $m$, and $K(m)$ is the
complete elliptic integral of the first kind. Equations (1) describe the
dynamics of a highly connected rotator (or hub), $x_{H}$, and $N-1$ linked
rotators (or leaves), $x_{i}$. The \textit{shape} parameter is taken to be
$m=0$ except for certain sets of rotators that are subjected to pulses of
variable width $\left(  m\in\left[  0,1\right]  \right)  $. The effect of
renormalization of the elliptic cosine argument is clear: with $T$ constant,
solely the pulse's wave form is varied by changing $m$ between $0$ and $1$.
Increasing $m$ makes the pulse narrower, and for $m\simeq1$ one recovers a
periodic sharply kicking excitation very close to the periodic $\delta$
function, but with finite amplitude and width (see Fig.~\ref{fig1} bottom). Also,
$\operatorname{cn}^{2}\left(  \Omega t;m=0\right)  =\cos^{2}\left(  \pi
t/T\right)  $, while at the other limit, $m=1$, the pulse area vanishes. It is
worth mentioning that the limiting case $m=\delta=\lambda=0$ corresponds to an
isolated Hamiltonian kicked rotator subjected to trigonometric pulses, which
has been used to describe the center-of-mass motion of cold atoms in an
amplitude-modulated standing wave of light [19], and that numerical studies
have shown the suppressory effectiveness of decreasing the impulse transmitted
by localized periodic pulses in homogeneous chains of DKRs [20].

Here we describe theoretical and numerical studies of the new chaos-control
scenario arising from Eq.~(1) by assuming parameter values such that each
isolated rotator driven by trigonometric pulses $\left(  m_{H}=m_{i}%
=0,i=1,...,N-1\right)  $ displays chaotic behavior characterized by a positive
Lyapunov exponent [21,22]. The remainder of the communication is organized as
follows. Section II studies both the chaotic dynamics and the oscillation
death (OD) [23] of isolated rotators (Eq.(1) with $\lambda=0$). Analytical
estimates of the chaotic threshold in parameter space are obtained by using
Melnikov's method (MM) [24,25], while the phenomenon of OD is anticipated
theoretically with the aid of an energy analysis. The interplay between
heterogeneous connectivity and local decrease of the pulse's impulse in
networks described by Eq.~(1) is discussed in Sec.~III. We characterize a
fairly complex regularization scenario, and determine how the effectiveness of
the local reshaping of pulses depends upon the number of control nodes and
their degree of connectivity. Finally, Sec.~IV is devoted to a discussion of
the major findings and to some concluding remarks.

\section{DYNAMICS OF ISOLATED ROTATORS}

Before considering the chaos-control scenario of DKRs coupled in a starlike
topology, it is necessary to understand the main features of the dynamics of
an isolated DKR:%
\begin{equation}
\overset{..}{x}+\operatorname{cn}^{2}\left(  \Omega t;m\right)  \sin
x=-\delta\overset{.}{x}. \tag{2}%
\end{equation}
In particular, we are interested in obtaining an analytical estimate of the
order-chaos threshold in parameter space, and in providing a theoretical
argument showing that the equilibrium $\left(  x=0,\overset{.}{x}=0\right)  $
may be a stable attractor of Eq.~(2) for pulses of a (certain) finite width
$\left(  m<1\right)  $. For the sake of clarity, we shall consider these
analyses separately.

\begin{figure}[tbp]
\begin{tabular}{c}
\includegraphics[width=6.5cm]{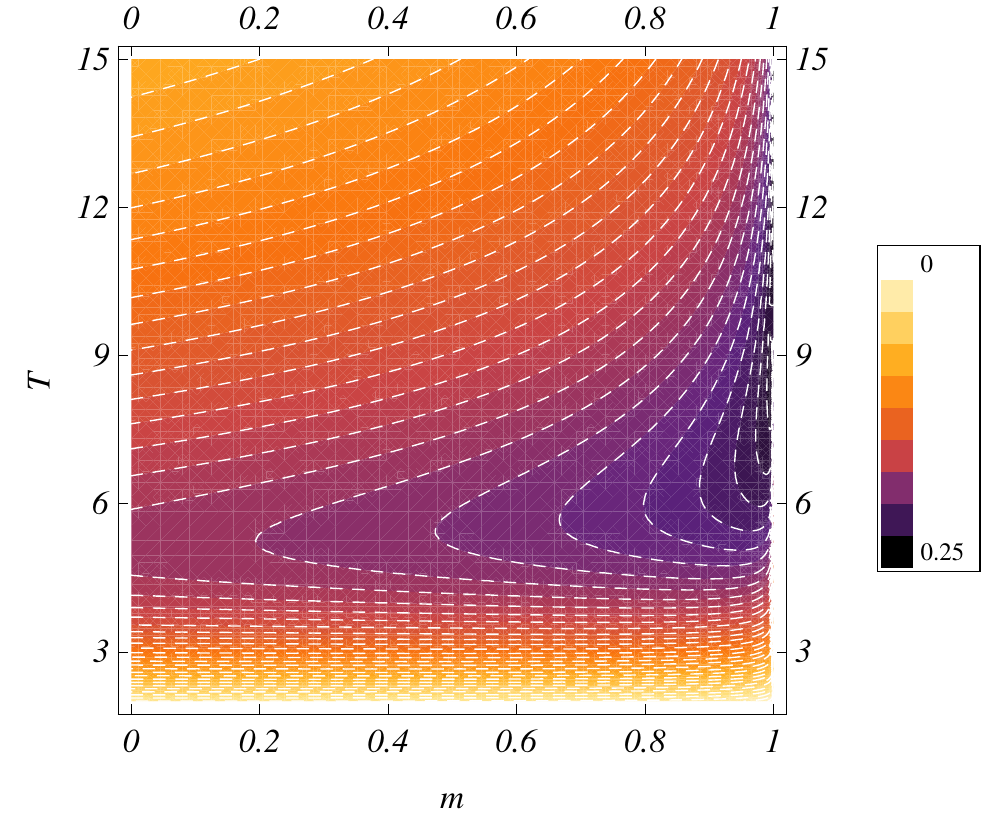} \\
\includegraphics[width=6.5cm]{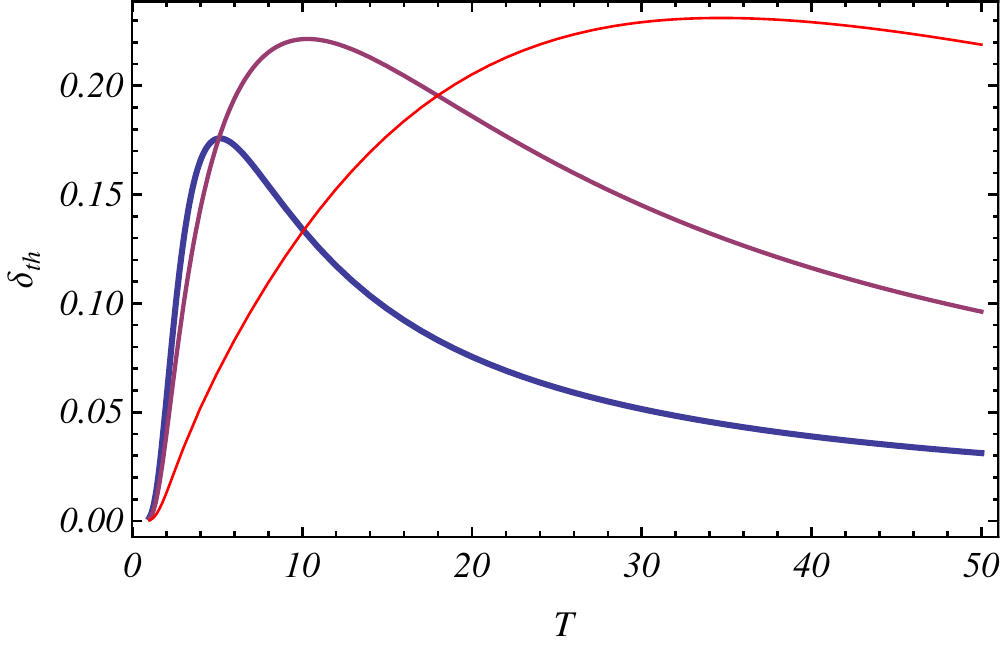} \\
\includegraphics[width=6.5cm]{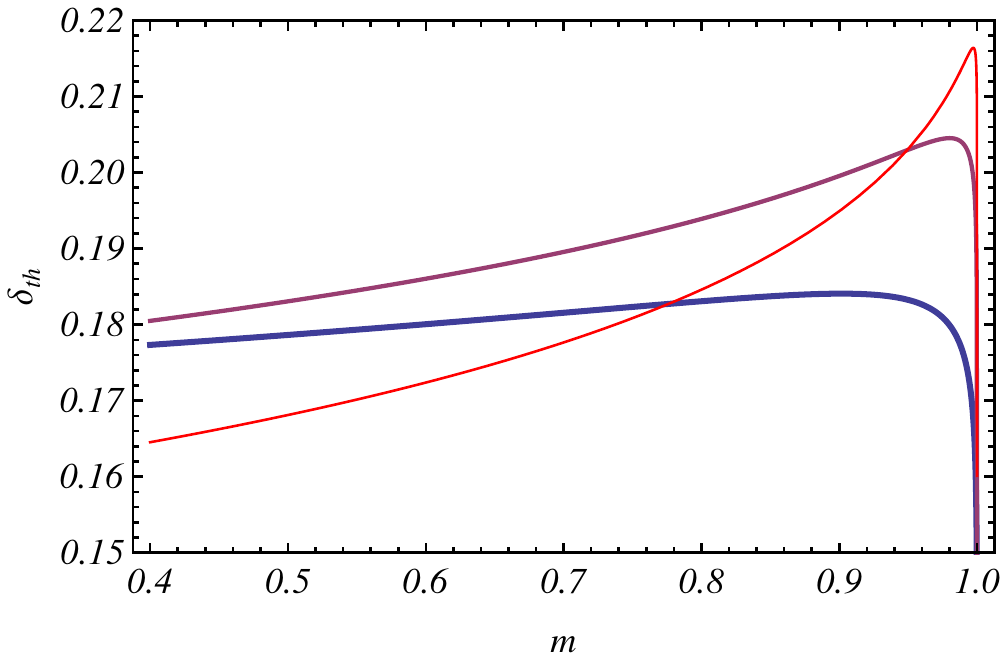} \\
\end{tabular}%
\caption{(Color online) Top: Contour plot of the chaotic threshold damping
$\delta_{th}\left(  T,m\right)  $ [cf. Eq.~(7)] vs shape parameter $m$ and
period $T$. Middle: Chaotic threshold damping $\delta_{th}\left(  T,m\right)
$ [cf. Eq.~(7)] vs period $T$ for $m=0$ (thick line), $m=0.999$ (medium line),
and $m=1-10^{-14}$ (thin line). Bottom: Chaotic threshold damping $\delta
_{th}\left(  T,m\right)  $ [cf. Eq.~(7)] vs period $m$ for $T=4.5$ (thick
line), $T=6$ (medium line), and $T=8$ (thin line). The quantities plotted are dimensionless.}
\label{fig2}
\end{figure}

\subsection{Order-chaos threshold}

\begin{figure}[tbp]
\begin{tabular}{c}
\includegraphics[width=6.5cm]{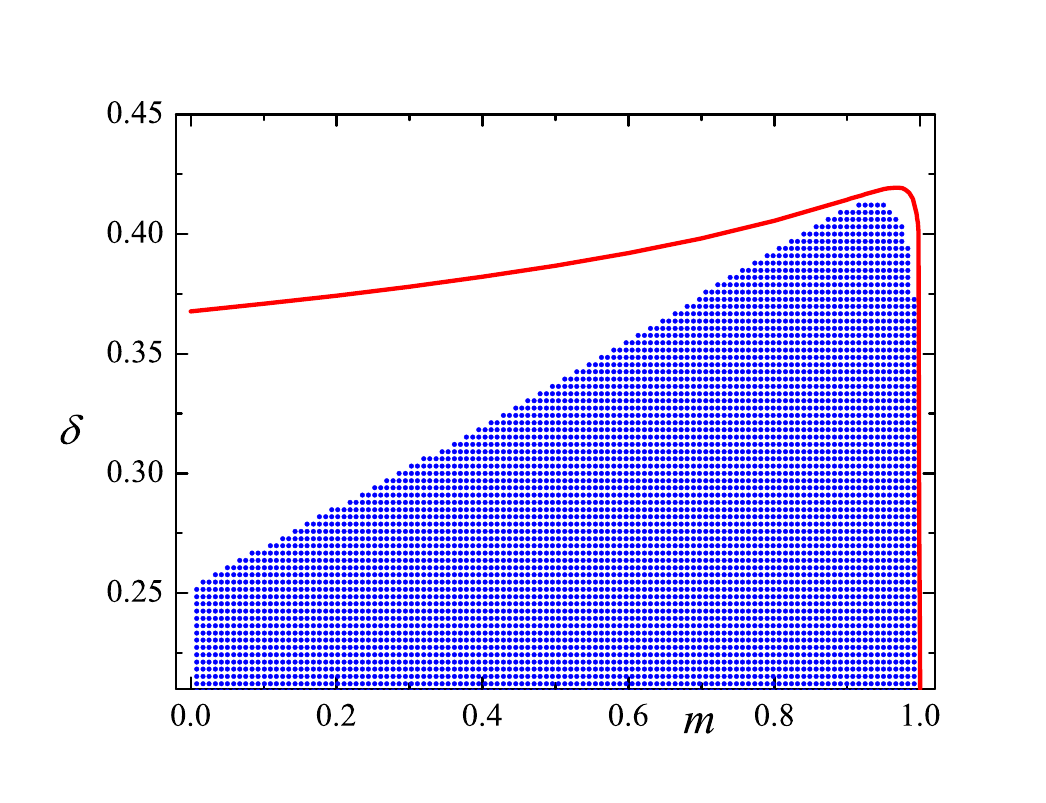}
\end{tabular}%
\caption{(Color online) Grid in the $m-\delta$ parameter plane for $T=5.52$.
Dots indicate that the respective leading LE is larger than $10^{-3}$. The
solid line denotes the theoretical estimate of the chaotic boundary
$2.1\delta_{th}\left(  T=5.52,m\right)  $ [cf. Eq. (7)] from MM.}
\label{fig3}
\end{figure}

To obtain analytical estimates of the chaotic threshold in parameter space
$\left(  T,m,\delta\right)  $, we first note that Eq.~(2) can be recast into
the form%
\begin{equation}
\overset{..}{x}+\sin x=-\delta\overset{.}{x}+\operatorname{sn}^{2}\left(
\Omega t;m\right)  \sin x,\tag{3}%
\end{equation}
where $\operatorname{sn}\left(  \cdot;m\right)  $ is the Jacobian elliptic
function of parameter $m$, and assume that the DKR (3) satisfies the MM
requirements, i.e., the dissipation and parametric excitation terms are
small-amplitude perturbations of the underlying conservative pendulum
$\overset{..}{x}+\sin x=0$. Melnikov introduced a function (the so-called
Melnikov function (MF), $M\left(  t_{0}\right)  $) which measures the distance
between the perturbed stable and unstable manifolds in the Poincar\'{e}
section at $t_{0}$. If the MF presents a simple zero, the manifolds intersect
transversally and chaotic instabilities result. See Refs.~[24,25] for more
details about MM. Regarding Eq.~(3), note that, in keeping with the assumption
of the MM [24,25], it is assumed that one can write $\delta=\varepsilon
\overline{\delta},$ where $0<\varepsilon\ll1$ while $\overline{\delta}$ is of
order unity. The term $\operatorname{sn}^{2}\left(  \Omega t;m\right)  \sin x$
in Eq.~(3) is not $O\left(  \varepsilon\right)  $, and one should not consider
it to be a perturbative term. However, this will be assumed in calculating the
MF so as to obtain an effective (qualitative) estimate of the chaotic
threshold in parameter space which may be useful in explaining the results of
the numerical experiments. Thus, bearing in mind this caveat, the application
of MM to Eq.~(3) yields the MF
\begin{align}
M^{\pm}\left(  t_{0}\right)   &  =-D+\sum_{n=1}^{\infty}a_{n}\left(  m\right)
b_{n}\left(  T\right)  \sin\left(  \frac{2n\pi t_{0}}{T}\right)  ,\nonumber\\
D &  \equiv8\delta,\nonumber\\
a_{n}\left(  m\right)   &  \equiv\frac{2n\pi^{3}}{mK^{2}(m)}%
\operatorname{csch}\left[  \frac{n\pi K(1-m)}{K(m)}\right]  ,\nonumber\\
b_{n}\left(  T\right)   &  \equiv\frac{4n^{2}\pi^{2}}{T^{2}}%
\operatorname{csch}\left(  \frac{n\pi^{2}}{T}\right)  ,\tag{4}%
\end{align}
where the positive (negative) sign refers to the top (bottom) homoclinic orbit
of the underlying conservative pendulum
\begin{align}
\theta_{0}\left(  t\right)   &  =\pm2\arctan\left[  \sinh\left(  t\right)
\right]  ,\nonumber\\
\overset{.}{\theta}_{0}\left(  t\right)   &  =\pm2\operatorname{sech}\left(
t\right)  .\tag{5}%
\end{align}
If $M^{\pm}\left(  t_{0}\right)  $ has a simple zero, then a heteroclinic
bifurcation occurs, signifying the onset of chaotic instabilities. From Eq.
(4) one sees that
\[
\sum_{n=1}^{\infty}a_{n}\left(  m\right)  b_{n}\left(  T\right)  \sin\left(
2n\pi t_{0}/T\right)  \leqslant\sum_{n=1}^{\infty}a_{n}\left(  m\right)
b_{n}\left(  T\right)  .
\]
If the damping coefficient is such that
\[
D\geqslant\sum_{n=1}^{\infty}a_{n}\left(  m\right)  b_{n}\left(  T\right)  ,
\]
this relationship represents a sufficient condition for $M^{\pm}\left(
t_{0}\right)  $ to always have the same sign, i.e., $M^{\pm}\left(
t_{0}\right)  \leqslant0$. Thus, a necessary condition for $M^{\pm}\left(
t_{0}\right)  $ to change sign at some $t_{0}$ is written
\begin{equation}
\delta<\delta_{th}\left(  T,m\right)  ,\tag{6}%
\end{equation}
where the chaotic threshold damping scales as%
\begin{equation}
\delta_{th}\left(  T,m\right)  \sim\frac{1}{8}\sum_{n=1}^{\infty}a_{n}\left(
m\right)  b_{n}\left(  T\right)  .\tag{7}%
\end{equation}
From Eq.~(7) one readily obtains
\[
\lim_{T\rightarrow0,\infty}\delta_{th}\left(  T,m\right)  =\lim_{m\rightarrow
1}\delta_{th}\left(  T,m\right)  =0,
\]
i.e., in such limits chaotic dynamics is not expected. Also, one finds that
$\delta_{th}\left(  T,m\right)  $ presents a maximum in the $m-T$ plane at
$\left(  m=m_{\max},T=T_{\max}\right)  $. A plot of $\delta_{th}\left(
T,m\right)  $ is shown in Fig.~\ref{fig2} top. Let us consider the chaotic threshold
damping as a function of $T$, holding $m$ constant. Plots of $\delta
_{th}\left(  T,m=const\right)  $ show that each curve presents a maximum
$T_{\max}=T_{\max}(m)$ such that $T_{\max}(m)$ increases from its value at
$m=0$ as $m\rightarrow1$ (see Fig.~\ref{fig2} middle). Now we study the chaotic
threshold damping as a function of $m$, holding $T$ constant. Plots of
$\delta_{th}\left(  T=const,m\right)  $ show that each curve presents a
maximum $m_{\max}=m_{\max}(T)$ such that $m_{\max}(T)$ increases as $T$ is
increased (see Fig 2 bottom). Thus, these MM-based predictions indicate that
by sufficiently decreasing the impulse transmitted by the pulses (time
integral over a period), i.e., when $m$ is sufficiently near $1$, is a
reliable procedure for suppressing chaos irrespective of the values of the
remaining parameters. Next, we compare the chaotic thresholds predicted from
MM and Lyapunov exponent (LE) calculations. It is worth mentioning that we
cannot expect too good a quantitative agreement between the two types of
results because MM is generally related with \textit{transient} chaos while LE
provides information concerning only steady motions. We compute LEs by using a
version of the algorithm introduced in Ref. [26]. We typically integrate up to
$10^{4}$ drive cycles for fixed period $T=5.52$. In a first step, we calculate
the leading LE for each point on a $100\times100$ grid, with shape parameter
$m$ and damping coefficient $\delta$ given by the horizontal and vertical
axes, respectively. Second, we construct the diagram shown in Fig. \ref{fig3} by only
plotting a point on the grid when the respective leading LE is larger than
$10^{-3}$. The chaotic threshold (solid line in Fig. \ref{fig3}) predicted from MM
gives a qualitative estimate for the upper boundary of the entire chaotic
region, as expected (recall the aforementioned caveats). Notwithstanding, the
theoretical estimate captures two main features of the numerically obtained
chaotic boundary: the existence of a maximum at a certain value of the shape
parameter, $m_{\max}^{LE}$, and that the chaotic boundary exhibits a
monotonously decreasing behavior as a function of the shape parameter from
$m=m_{\max}^{LE}$. Note that the corresponding theoretically predicted maximum
$m_{\max}=m_{\max}(T=5.52)$ is very close to $m_{\max}^{LE}$.

We will show in Sec.~III how numerical simulations of starlike networks of
DKRs confirmed the effectiveness of this chaos-control procedure.

\subsection{Energy-based analysis}

\begin{figure}[tbp]
\begin{tabular}{c}
\includegraphics[width=6.5cm]{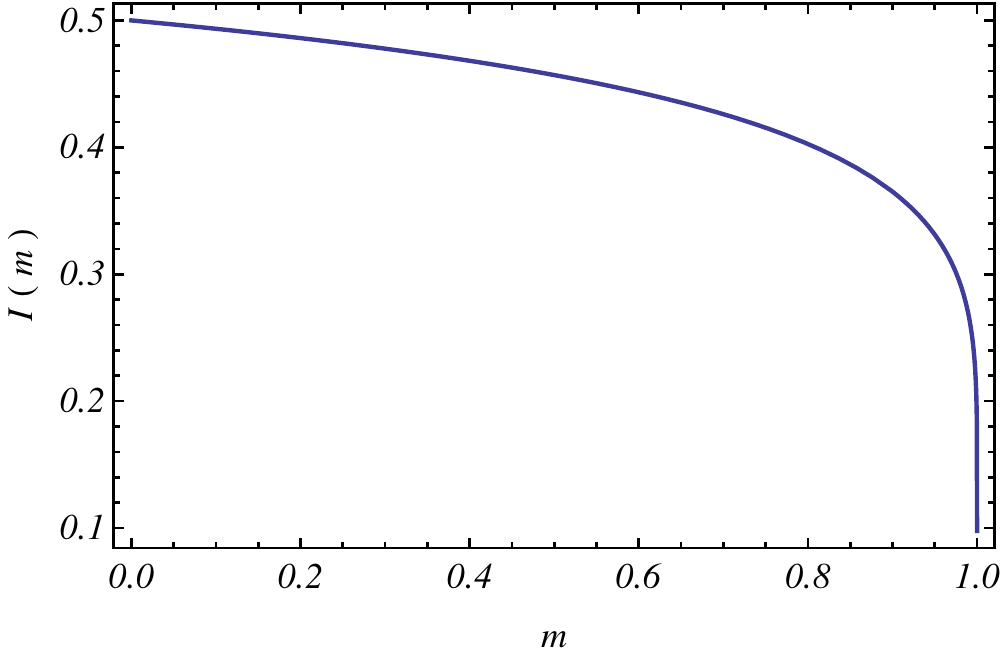}
\end{tabular}%
\caption{(Color online) Impulse function $I\left(  m\right)  $ versus shape
parameter $m$ (cf. Eq.~(11)). The quantities plotted are dimensionless.}
\label{fig4}
\end{figure}

By analyzing the variation of the system's kinetic energy, one
straightforwardly predicts the occurrence of the phenomenon of OD. Indeed,
Eq.~(3) has the associated energy equation
\begin{equation}
\frac{dE_{K}}{dt}=-\delta\overset{.}{x}^{2}-\operatorname{cn}^{2}\left(
\Omega t;m\right)  \overset{.}{x}\sin x, \tag{8}%
\end{equation}
where $E_{K}(t)\equiv\left(  1/2\right)  \overset{.}{x}^{2}\left(  t\right)  $
is the kinetic energy function. Integration of Eq.~(8) over \textit{any}
interval $\left[  nT,nT+T\right]  $, $n=0,1,2,...$, yields
\begin{align}
E_{K}\left(  nT+T\right)   &  =E_{K}(nT)-\delta\int_{nT}^{nT+T}\overset{.}%
{x}^{2}\left(  t\right)  dt\nonumber\\
&  -\int_{nT}^{nT+T}\operatorname{cn}^{2}\left(  \Omega
t;m\right)  \dot{x}\left(  t\right)  \sin x(t)dt. \tag{9}%
\end{align}
Now, after applying \ the first mean value theorem for integrals [28] together
with well-known properties of the Jacobian elliptic functions [27] to the last
two integrals on the right-hand side of Eq.~(9), one has
\begin{align}
E_{K}\left(  nT+T\right)   &  =E_{K}(nT)-\delta\overset{.}{x}^{2}\left(
t^{\ast\ast}\right)  T\nonumber\\
&  -\overset{.}{x}\left(  t^{\ast}\right)  \sin\left[  x\left(  t^{\ast
}\right)  \right]  I\left(  m\right)  T, \tag{10}%
\end{align}
where $t^{\ast},t^{\ast\ast}\in\left[  nT,nT+T\right]  $, while
\begin{equation}
I\left(  m\right)  \equiv\frac{E(m)+(m-1)K(m)}{mK(m)} \tag{11}%
\end{equation}
is the \textit{impulse} transmitted over a period $T=1$ with $E(m)$ being the
complete elliptic integral of the second kind. From Eq.~(11) one
straightforwardly obtains $I\left(  m=0\right)  =1/2,I\left(  m=1\right)  =0$.
A plot of $I\left(  m\right)  $ is shown in Fig.~\ref{fig4}. Now, if we consider fixing
the parameters $\left(  \delta,T\right)  $ for the DKR to exhibit chaotic
dynamics at $m=0$, there always exists an $n=n^{\ast}$ such that the kinetic
energy increment
\[
\Delta E_{K}^{m=0}\equiv E_{K}\left(  n^{\ast}T+T\right)  -E_{K}(n^{\ast
}T)>0.
\]

\begin{figure}[tbp]
\begin{tabular}{c}
\includegraphics[width=6.5cm]{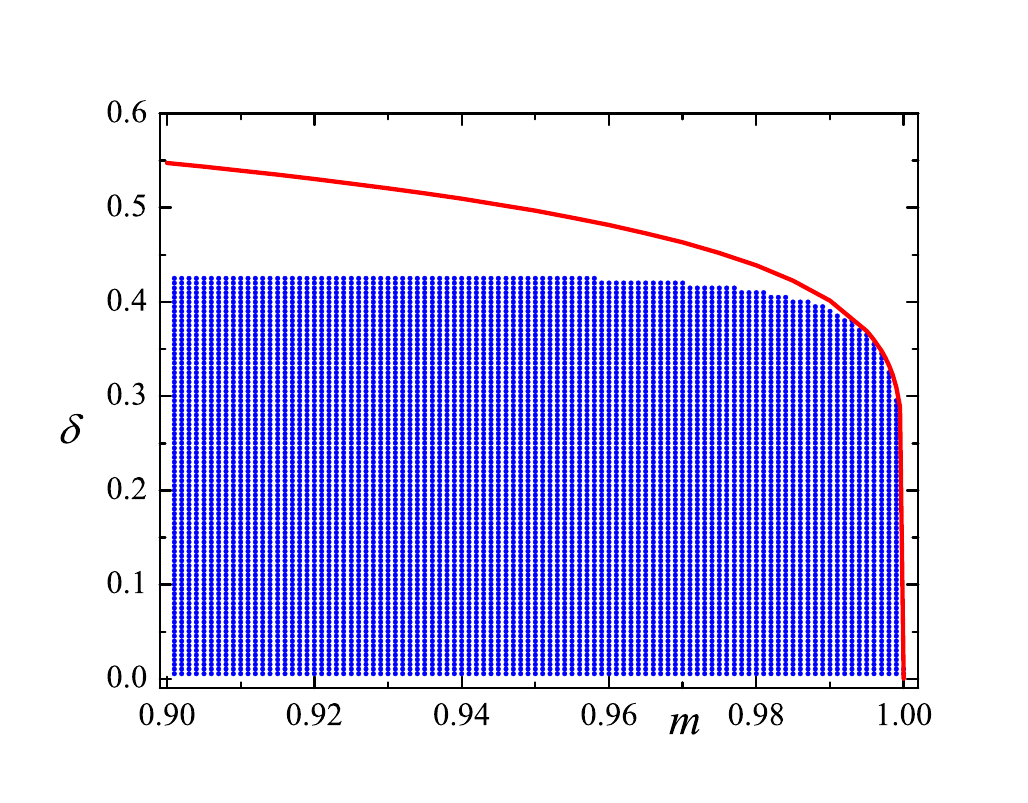}
\end{tabular}%
\caption{(Color online) Stability boundary of the equilibrium $\left(
x=0,\overset{.}{x}=0\right)  $ in the $m-\delta$ parameter plane for $T=5.52$.
The instability region (dots) was numerically calculated on a grid of
$100\times100$ points. The solid line denotes the theoretical estimate of the
stability boundary $\delta_{c}=1.5I(m)$ from Eq. (12). The quantities plotted
are dimensionless.}
\label{fig5}
\end{figure}

In this situation, one decreases the impulse by increasing the shape parameter
from $m=0$ while holding the remaining parameters constant. Equations (10) and
(11) predict that, for each $n^{\ast}$, there always exists a minimum critical
value $m=m_{c}>0$ such that the corresponding energy increment $\Delta
E_{K}^{m=m_{c}}<0$ for \textit{all} $n>n^{\ast}$, and hence the equilibrium
$\left(  x=0,\overset{.}{x}=0\right)  $ is the single attractor of the DKR for
$m\geqslant m_{c}$. Note that this property comes ultimately from the behavior
of the impulse $I\left(  m\right)  $ as the shape parameter $m\rightarrow1$
(see Fig.~\ref{fig4}), i.e., that the DKR effectively behaves as a purely damped
pendulum for sufficiently narrow pulses. Thus, one straightforwardly obtains
from Eqs. (10) and (11) that, for sufficiently narrow pulses, the equilibrium
$\left(  x=0,\overset{.}{x}=0\right)  $ is the single attractor of the DKR
when $\delta>\delta_{c}$, where the critical damping coefficient scales as%
\begin{equation}
\delta_{c}\sim I(m). \tag{12}%
\end{equation}
Numerical experiments confirmed the validity of this scaling for sufficiently
narrow pulses, as in the example shown in Fig. \ref{fig5}.

\section{LOCALIZED CONTROL IN STARLIKE NETWORKS}

\begin{figure}[tbp]
\begin{tabular}{c}
\includegraphics[width=6.5cm]{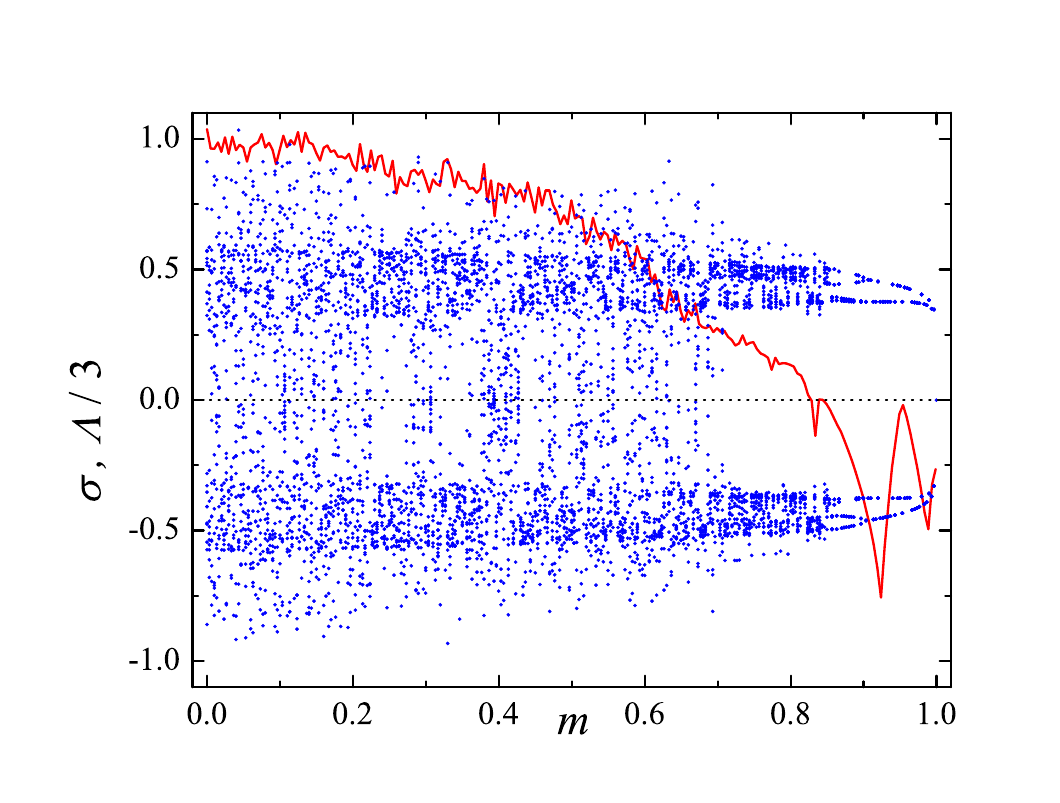}
\end{tabular}%
\caption{(Color online) Bifurcation diagrams of the average velocity $\sigma$
(blue (black) dots) and leading LE $\Lambda$ (red (gray) line) as a function
of the shape parameter $m=m_{j}$ for the case of a central rotator subjected
to trigonometric pulses $\left(  m_{H}=0\right)  $, $N=10,\lambda
=0.1,\delta=0.2,T=5.52$, and $M=5$ peripheral rotators $x_{j}$ subjected to
impulse control. The quantities plotted are dimensionless.}
\label{fig6}
\end{figure}

In this section, we study the relative effectiveness of locally reshaping the
pulses $\operatorname{cn}^{2}\left(  \Omega t;m\right)  $, in the sense of
decreasing their impulse, on $M$ nodes of chaotic starlike networks of $N$
DKRs (cf. Eq.~(1), $M<N$) while holding the remaining parameters constant.
Before applying any control, we assume parameter values $\left(
\delta,T\right)  $ such that each isolated rotator driven by trigonometric
pulses $\left(  m_{H}=m_{i}=0,i=1,...,N-1\right)  $ displays chaotic behavior.
Equation (1) was numerically integrated using a fourth-order Runge-Kutta
algorithm with a time step $\Delta t=0.001$. To visualize the global spatiotemporal
dynamics of networks, we calculated the average velocity%
\begin{equation}
\sigma\left(  jT\right)  \equiv\frac{1}{N}\sum_{i=1}^{N}\frac{dx_{i}}%
{dt}\left(  jT\right)  , \tag{13}%
\end{equation}
where $j$ is an integer multiple of the pulse period $T$, while the degree of
synchronization is characterized by the correlation function%
\begin{equation}
C\equiv\frac{2}{N(N-1)}\sum_{\left(  il\right)  }\left\langle \cos\left(
x_{i}-x_{l}\right)  \right\rangle _{t}, \tag{14}%
\end{equation}
with the summation being over all pairs of rotators, and where $\left\langle
\cdot\right\rangle _{t}$ indicates time averaging over a predefined
(sufficiently long) observation window. Note that $C$ is $1(0)$ for the
perfectly synchronized (desynchronized) state. Calculations of LEs for the
starlike networks [Eq. (1)] confirmed the reliability of the information
provided by bifurcation diagrams of the average velocity $\sigma$ concerning
transitions order-chaos. An illustrative example is shown in Fig. \ref{fig6} for the
case $N=10,M=5,\lambda=0.1,\delta=0.2,T=5.52$.

\subsection{Control on a single peripheral rotator}

\begin{figure}[tbp]
\begin{tabular}{c}
\includegraphics[width=6.5cm]{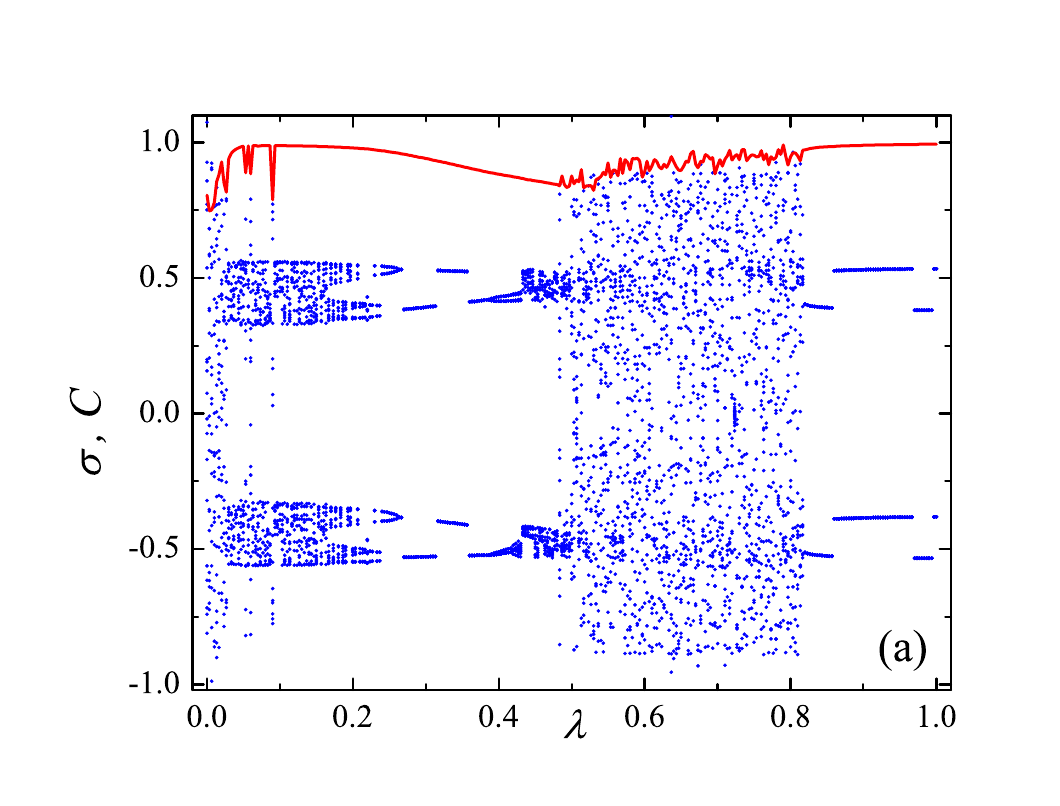} \\
\includegraphics[width=6.5cm]{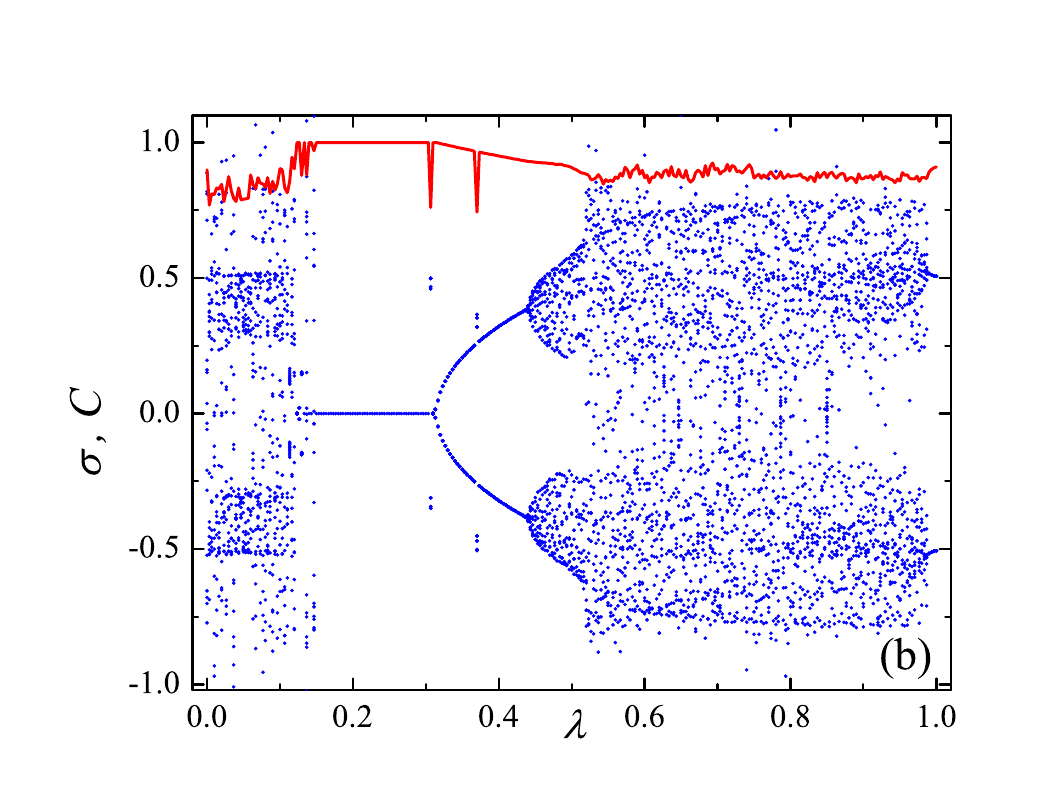}
\end{tabular}%
\caption{(Color online) Bifurcation diagrams of the average velocity $\sigma$
(blue (black) dots) and correlation function $C$ (red (gray) line) as a
function of the coupling $\lambda$ for the case of a central rotator subjected
to trigonometric pulses, $N=10,M=1,\delta=0.2,T=5.52$, and two values of the
shape parameter: (a) $m_{j}=0.9$ and (b) $m_{j}=1-10^{-14}$. The quantities
plotted are dimensionless.}
\label{fig7}
\end{figure}

\begin{figure}[tbp]
\begin{tabular}{c}
\includegraphics[width=6.5cm]{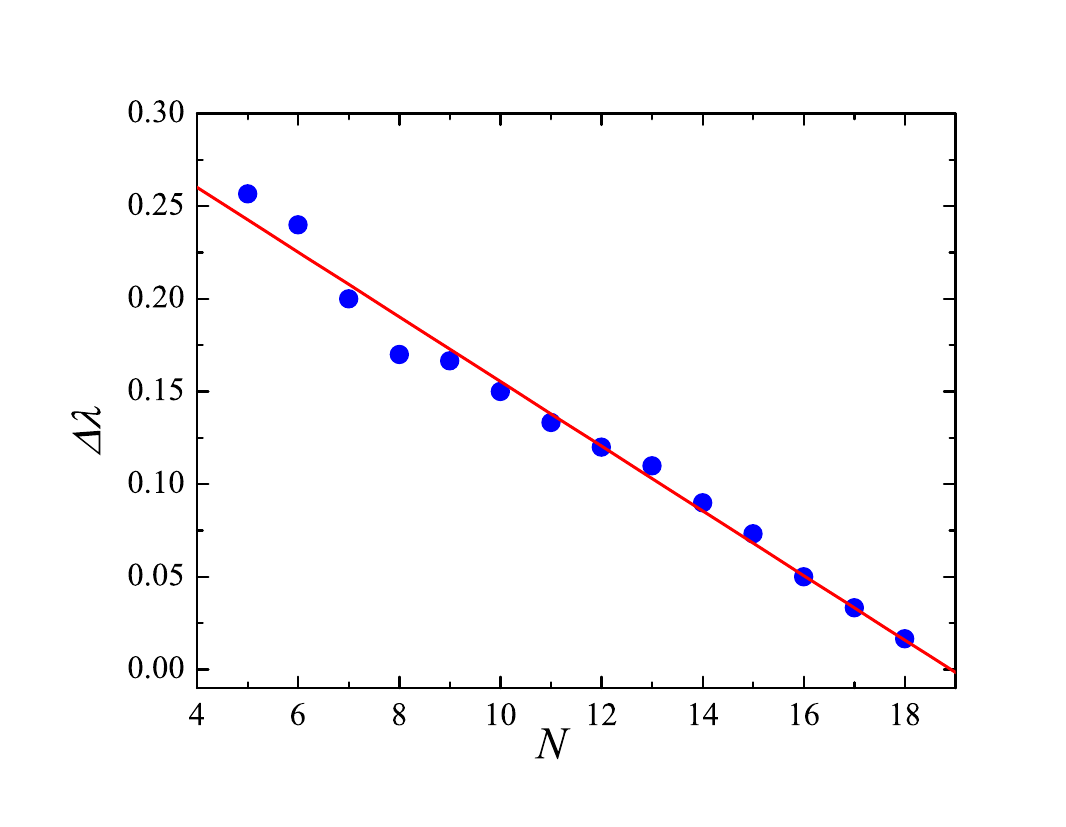}
\end{tabular}%
\caption{(Color online) Width of the coupling interval where OD occurs,
$\Delta\lambda\equiv\lambda_{\max}-\lambda_{\min}$ (see the text), as a
function of the number of rotators $N$ for $M=1,\delta=0.2,T=5.52,m_{j}%
=1-10^{-14}$. The line denotes the linear fit $\left(
0.32988-0.01744N\right)  $. The quantities plotted are dimensionless.}
\label{fig8}
\end{figure}

\begin{figure}[tbp]
\begin{tabular}{cc}
\includegraphics[width=4cm]{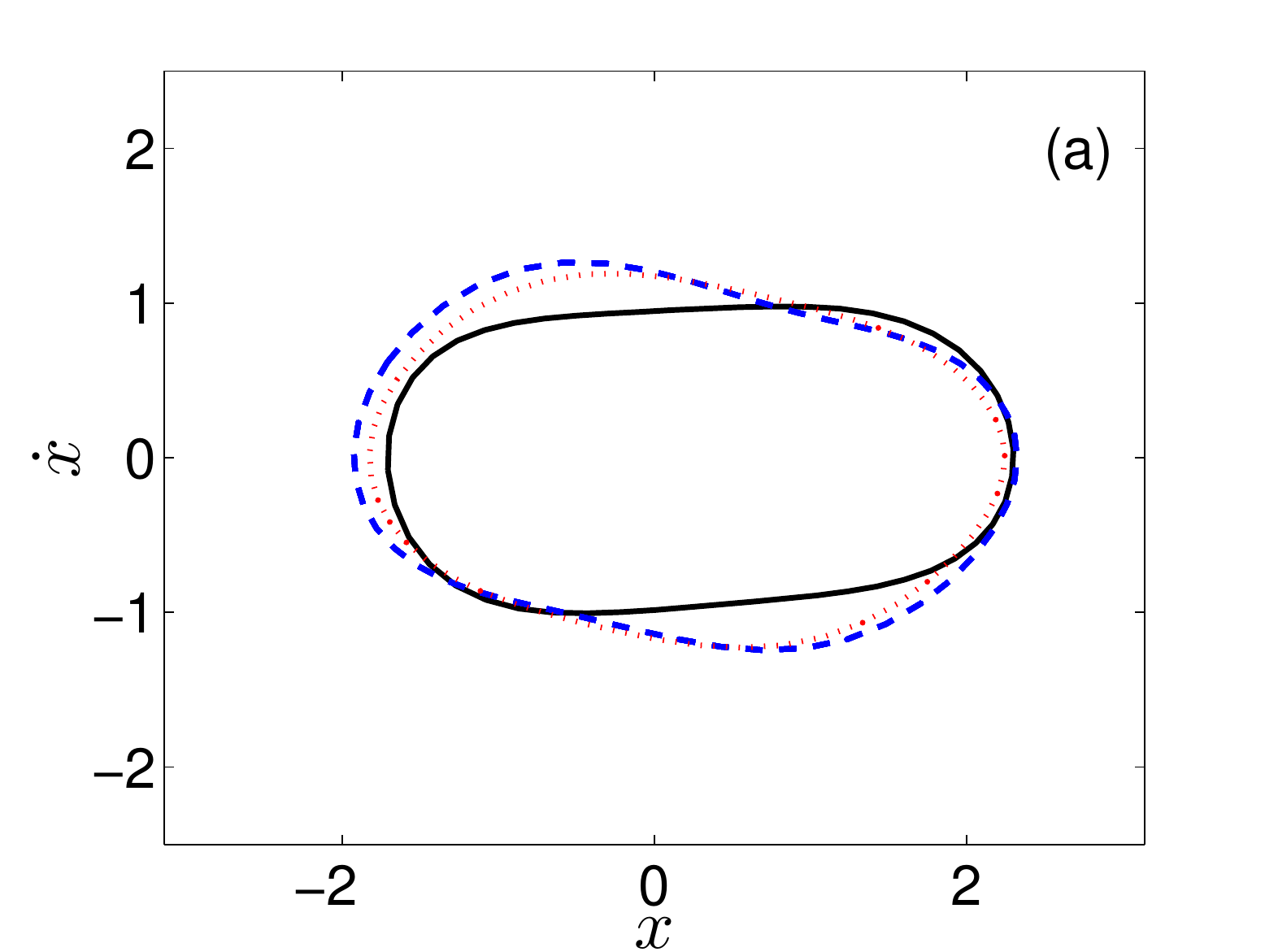} &
\includegraphics[width=4cm]{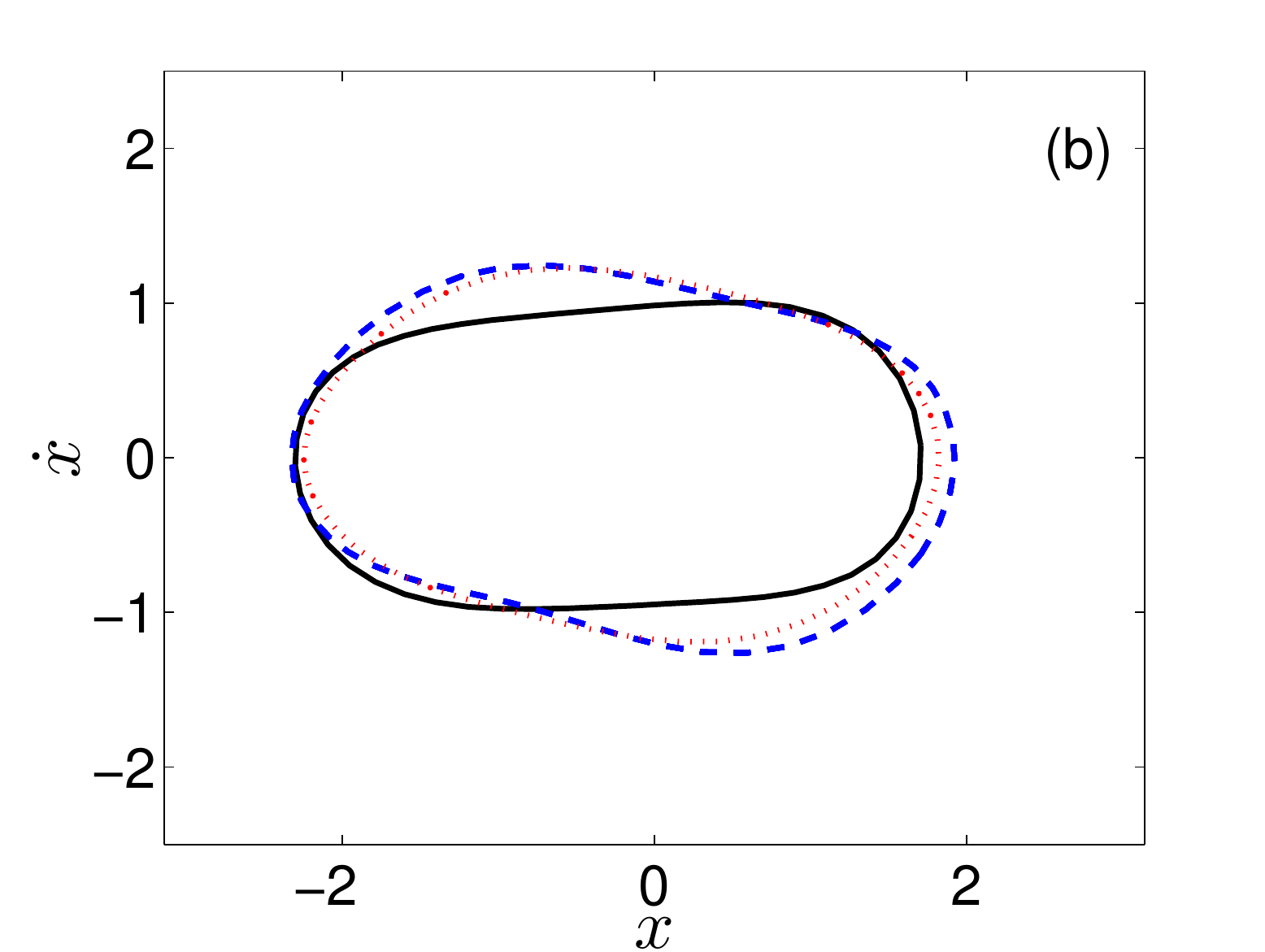} \\
\includegraphics[width=4cm]{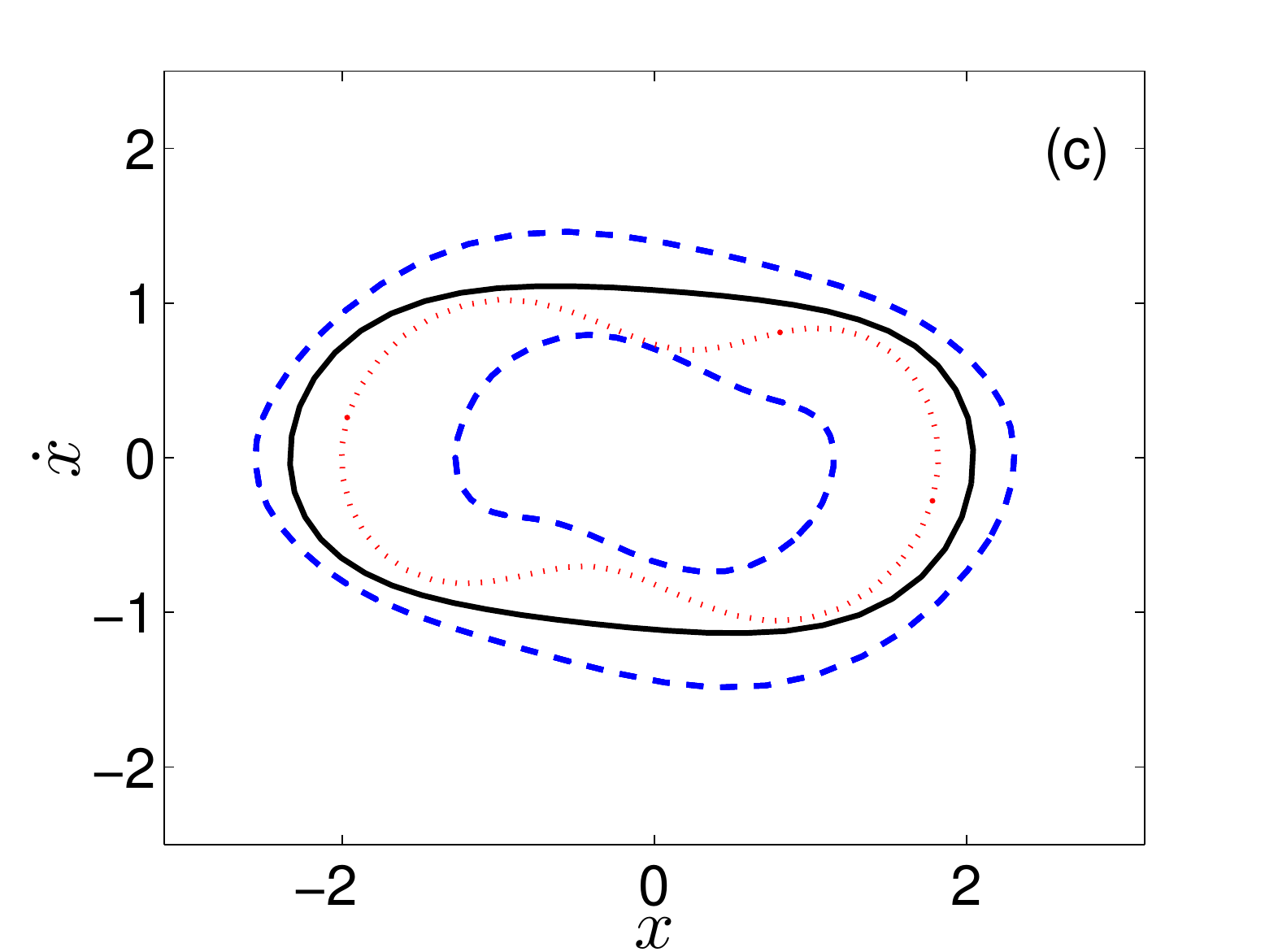} &
\includegraphics[width=4cm]{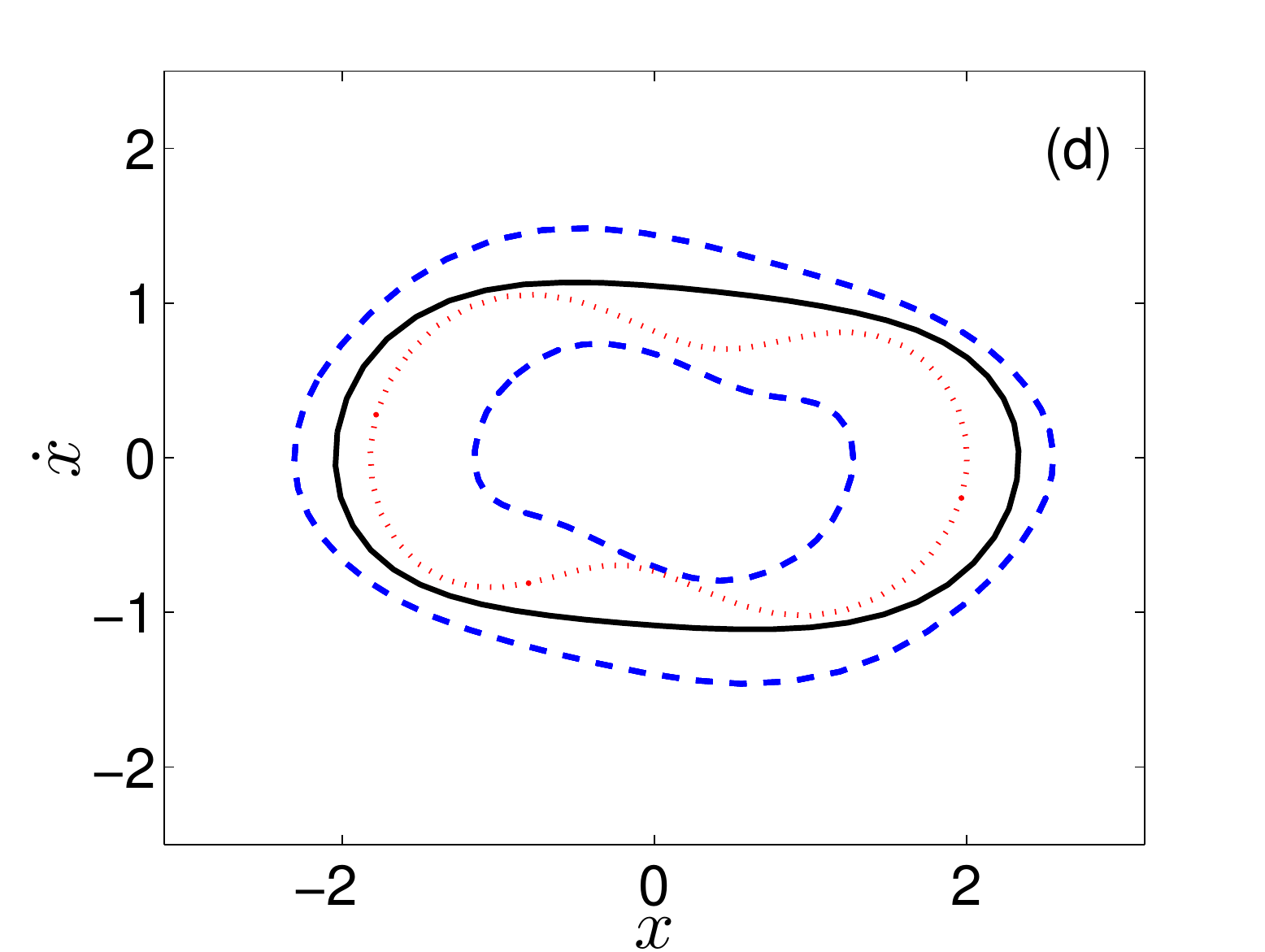} \\
\end{tabular}%
\caption{(Color online) Phase-space portraits of different periodic attractors
corresponding to a starlike network [Eq. (1)] within a range of multistability
for $N=10,M=1,\delta=0.2,T=5.52,\lambda=0.3$. (a) Hub ($m_{H}=0$, dotted
line), single leaf subjected to control ($m_{1}=0.9$, solid line), and
remaining leaves being fully synchronized ($m_{i}=0,i=2,...,9$, dashed line).
(b) Antisymmetric versions of the attractors plotted in (a), respectively. (c)
Hub ($m_{H}=0$, dotted line), single leaf subjected to control ($m_{1}=0.9$,
solid line), and remaining leaves grouped into two clusters of distinct
synchronized dynamics (dashed lines). (d) Antisymmetric versions
of the attractors plotted in (c), respectively.}
\label{fig9}
\end{figure}

Let us first consider the effect of decreasing the pulses' impulse on a single
peripheral rotator $x_{j}$ $\left(  M=1\right)  $ while the remaining
rotators, including the hub, are subjected to trigonometric pulses $\left(
m_{H}=m_{i}=0,i=1,...N-1,i\neq j\right)  $. Note that this could be, \textit{a
priori}, the most unfavorable case in terms of completely regularizing the
whole network. Numerical simulations indicate, however, that regularization to
periodic states is possible over certain coupling intervals even for
relatively wide pulses, such as for $m_{j}=0.9$ (see Fig.~\ref{fig7}(a)), as expected
from the above MM-based predictions. The symmetry of the bifurcation diagrams
comes from the DKR's symmetry with respect to the transformation $\left(
x_{i}\rightarrow-x_{i}\right)  $, i.e., if $\left[  x_{i}(t),\overset{.}%
{x}_{i}\left(  t\right)  \right]  $ is a solution of Eq.~(1), then so is
$\left[  -x_{i}(t),-\overset{.}{x}_{i}\left(  t\right)  \right]  $. The
bifurcation diagram was constructed by means of a Poincar\'{e} map at the
parameters indicated in the caption to Fig.~\ref{fig7}. Starting at $\lambda=0$, and
taking the transient time as $500$ pulse periods after every increment of
$\Delta\lambda=3.3\times10^{-3}$, we sampled $20$ pulse periods by picking up
the first $\sigma$ values of every pulse cycle, while to obtain the
correlation function [see Eq. (14)] we calculated $C$ averaged over $200$
additional pulse periods. In accordance with the above energy analysis, we
typically find that the phenomenon of OD occurs over certain coupling
intervals for sufficiently narrow pulses, the equilibrium $\left(
x=0,\overset{.}{x}=0\right)  $ being the asymptotic behavior of the perfectly
synchronized network, as in the illustrative instance shown in Fig.~\ref{fig7}(b) for
$m_{j}=1-10^{-14}$. In general, the equilibrium $\left(  x=0,\overset{.}%
{x}=0\right)  $ becomes stable at a certain value $\lambda=\lambda_{\min}$ via
a boundary crisis, while it becomes unstable at a certain higher value
$\lambda=\lambda_{\max}$ via a supercritical Hopf bifurcation [29]. These
threshold values of the coupling, $\lambda_{\max,\min}$, depend upon the
remaining parameters. In particular, the dependence on the number of rotators
$N$ of the width of the coupling interval in which OD occurs, $\Delta
\lambda=\Delta\lambda\left(  N\right)  \equiv\lambda_{\max}\left(  N\right)
-\lambda_{\min}\left(  N\right)  $, follows a \textit{linear} law, as is shown
in Fig.~\ref{fig8}. Note that $\Delta\lambda\rightarrow0$ as $N$ approximates a
sufficiently large but finite number of rotators. Besides the correlation
function $C$, bifurcation diagrams of the average velocity $\sigma$ also
provide useful information regarding the existence of multistability [30] over
certain ranges of parameters. In the present case of starlike networks which
are sufficiently far from the Hamiltonian limiting case, multistability comes
from the conjoint effect of localized-control-induced heterogeneity and the
aforementioned parity symmetry. Typically, we found that the ranges of
existence of particular attractors are relatively narrow so that the
qualitative behavior of the starlike networks can change dramatically after
slightly varying their parameters. An example of multistability of periodic
attractors is found to occur over a short coupling interval around
$\lambda=0.3$ for the fixed parameters $N=10,M=1,\delta=0.2,T=5.52,m=0.9$ (see
Fig. \ref{fig7}(a)). Thus, after exploring the initial conditions space, one finds many
coexisting periodic attractors which correspond to pairs of antisymmetric
$2T$-periodic attractors (see Fig. \ref{fig9}). Other different cases, including the
coexistence of periodic and chaotic attractors, were detected: An exhaustive
study of multistability is beyond the scope of the present work.

\subsection{Control on an increasing number of peripheral rotators}

\begin{figure}[tbp]
\begin{tabular}{cc}
\includegraphics[width=4.25cm]{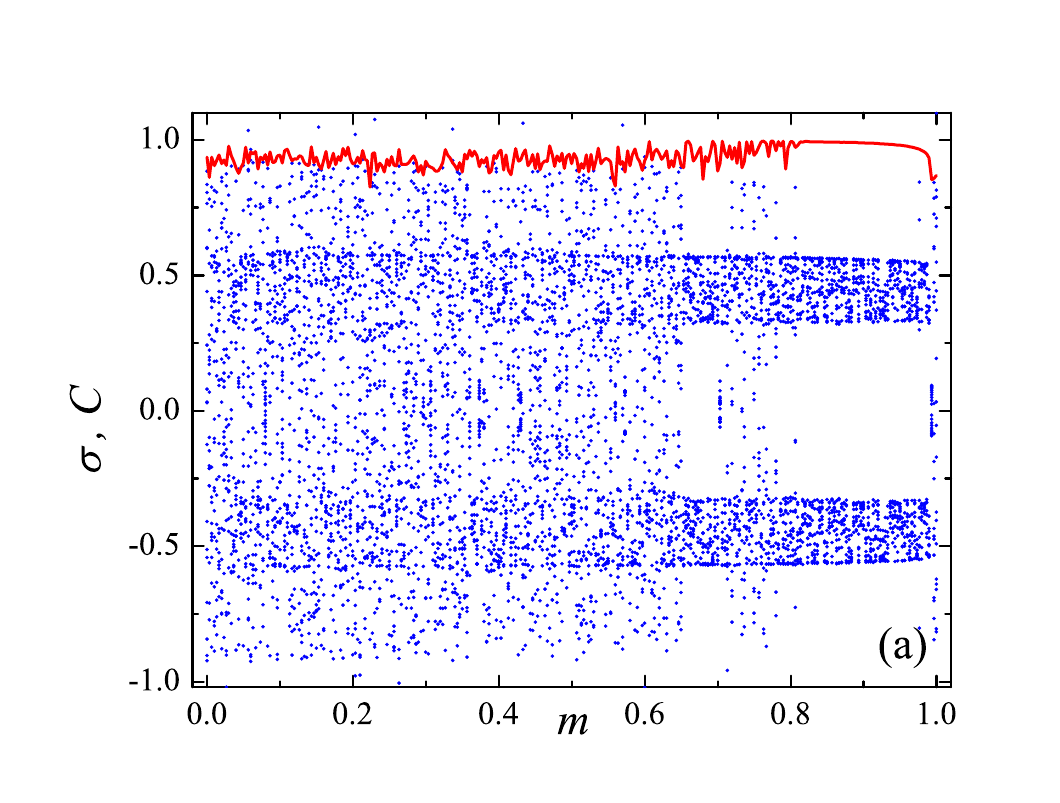} &
\includegraphics[width=4.25cm]{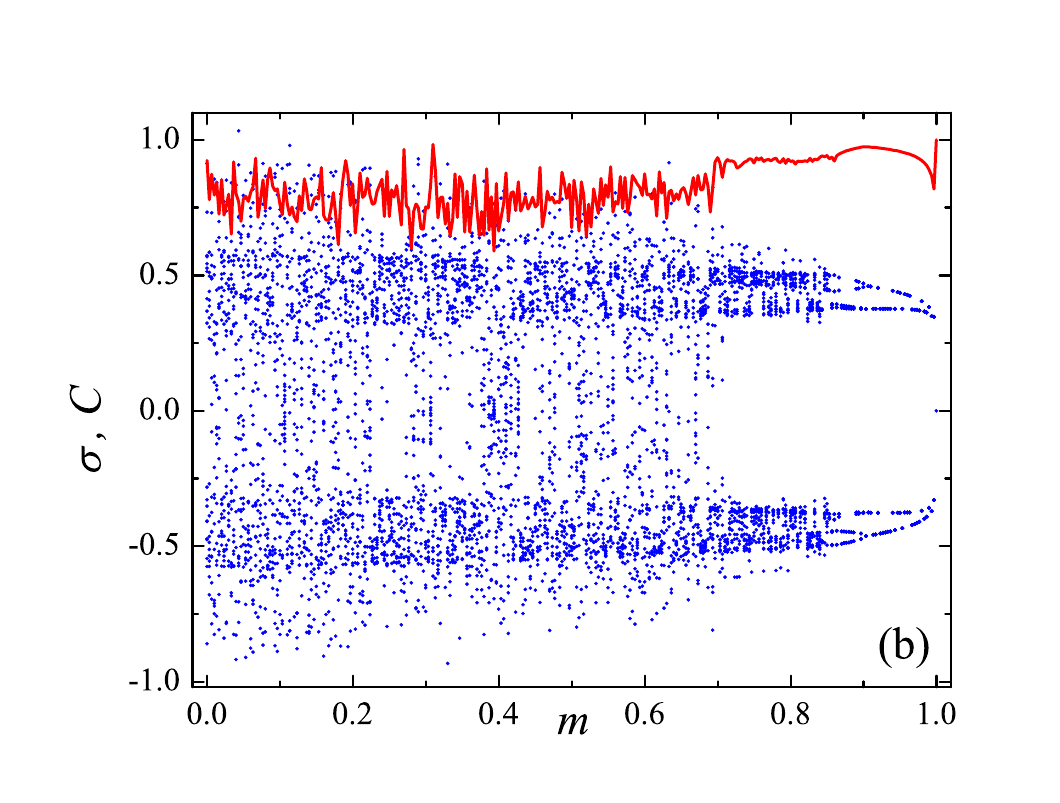} \\
\includegraphics[width=4.25cm]{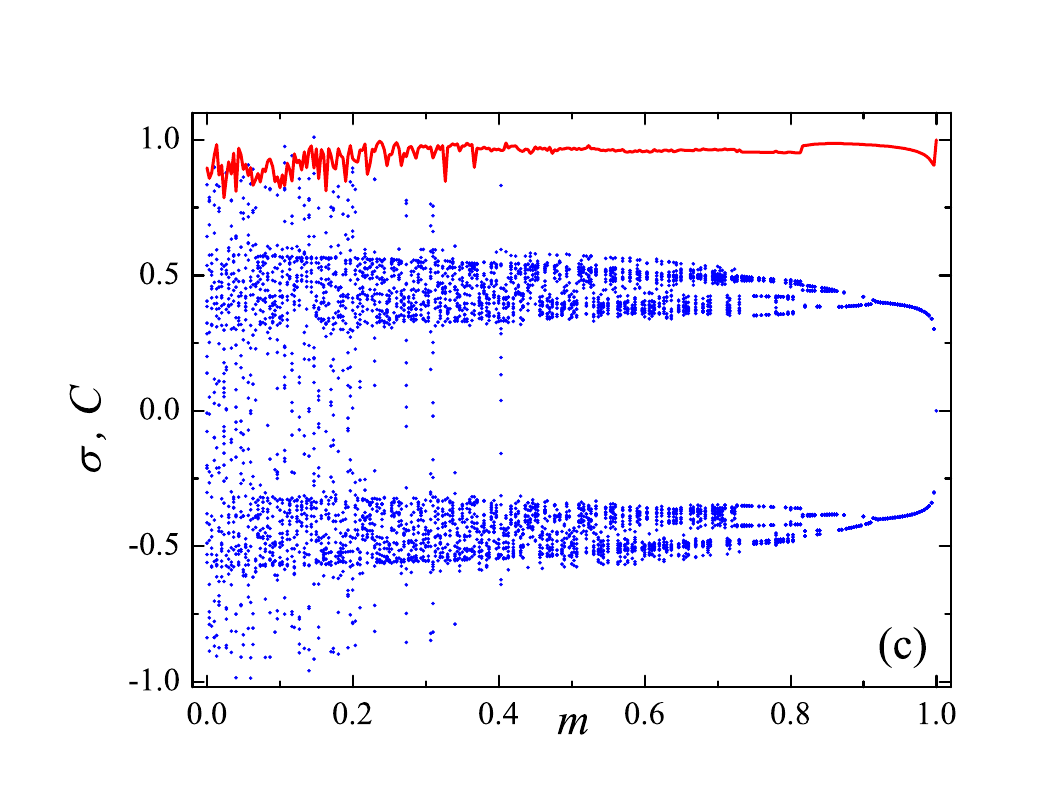} &
\includegraphics[width=4.25cm]{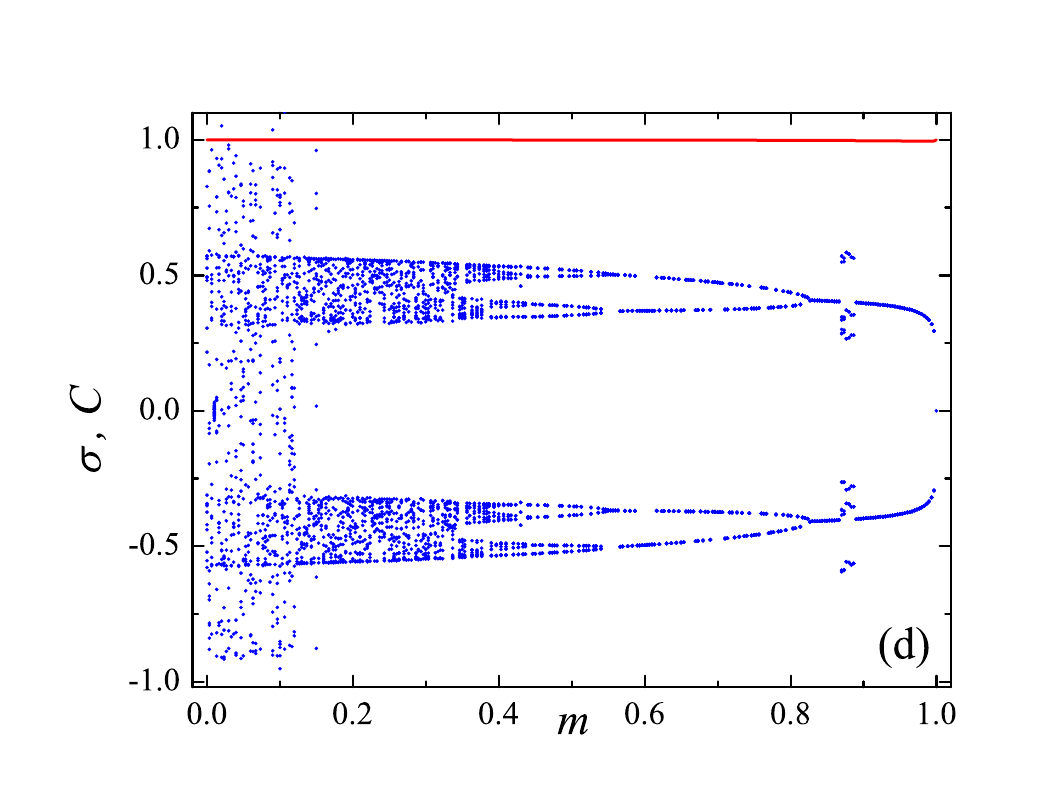} \\
\end{tabular}%
\caption{(Color online) Bifurcation diagrams of the average velocity $\sigma$
(blue (black) dots) and correlation function $C$ (red (gray) line) as a
function of the shape parameter $m=m_{j}$ for the case of a central rotator
subjected to trigonometric pulses $\left(  m_{H}=0\right)  $, $N=10,\lambda
=0.1,\delta=0.2,T=5.52$, and four values of the number of peripheral rotators
$x_{j}$ subjected to impulse control: (a) $M=1$, (b) $M=5$, (c) $M=7$, and (d)
$M=9$. The quantities plotted are dimensionless.}
\label{fig10}
\end{figure}

It is interesting to study the accumulative effect of decreasing the pulses'
impulse on an increasing number of peripheral rotators $\left(  M>1\right)  $,
while the hub remains subjected to trigonometric pulses, in the weak coupling
regime where synchronization phenomena do not yet dominate the networks'
dynamics. We start from a situation where regularization is not possible for
almost any value of the shape parameter when the pulses' impulse is decreased
on a single peripheral rotator ($M=1$, see Fig.~\ref{fig10}(a)). Then, by increasing
$M$ from unity, one typically obtains regularization of the whole network for
\textit{sufficiently} narrow pulses, on the one hand, and a deterioration of
the synchronization of the chaotic dynamics for wider pulses on the other (see
Fig.~\ref{fig10}(b)). This further effect, which is especially noticeable when $M$ is
near $N/2$ (compare the case $M=5$ with the cases $M=1$ and $M=7$, cf.
Figs.~\ref{fig10}(b), \ref{fig10}(a), \ref{fig10}(c), respectively), is due to the appearance of two
different synchronized populations of rotators subjected respectively to
pulses of different widths. As $M$ approximates $N-1$, i.e., when the impulse
control is applied to all the peripheral rotators, the width of the interval
$\Delta m$ where reshaping-induced regularization occurs increases, while the
synchronization increases drastically even when the dynamics is chaotic, as
for $M=9=N-1$ (see Fig.~\ref{fig10}(d)).

\subsection{Control on the central rotator}

\begin{figure}[tbp]
\begin{tabular}{c}
\includegraphics[width=6.5cm]{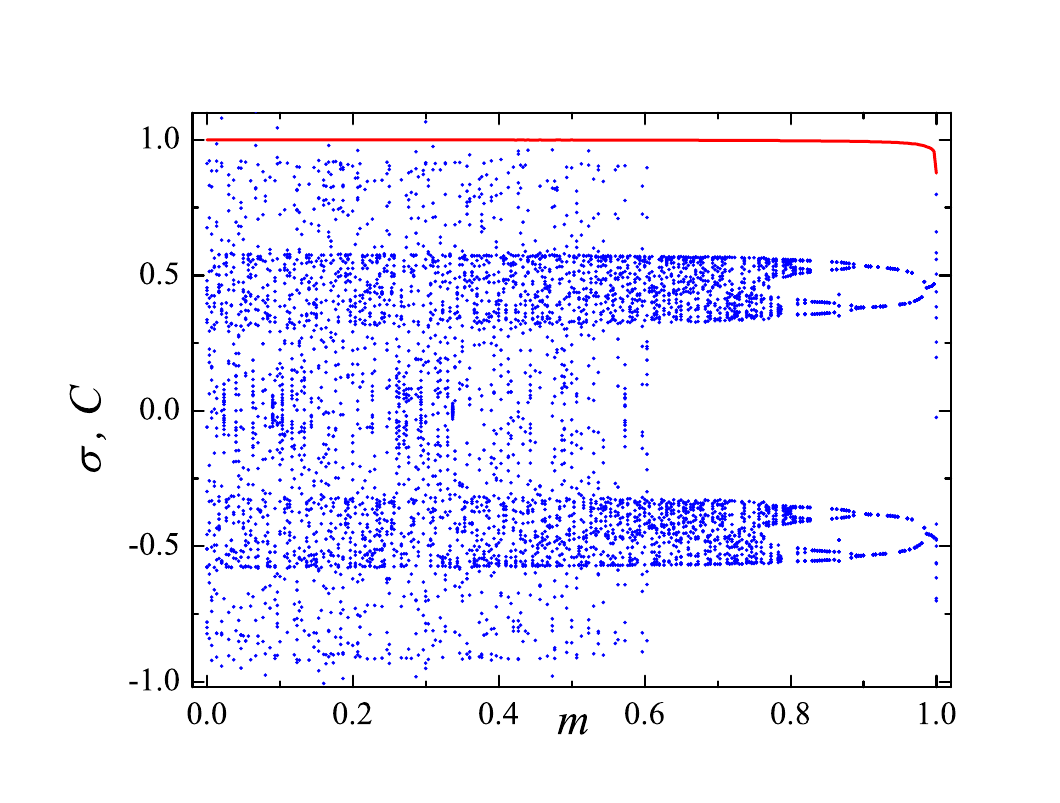}
\end{tabular}%
\caption{(Color online) Bifurcation diagrams of the average velocity $\sigma$
(blue (black) dots) and correlation function $C$ (red (gray) line) as a
function of the shape parameter $m=m_{H}$ when the central rotator (hub) is
the single rotator subjected to impulse control for $N=10,\lambda
=0.1,\delta=0.2,T=5.52$. The quantities plotted are dimensionless.}
\label{fig11}
\end{figure}

In the present subsection and the next, we study the role played by the degree
of connectivity in the reshaping-induced chaos-control scenario by decreasing
the pulses' impulse on the central rotator. In the case of a single control
(the present subsection), one finds that controlling the most highly connected
rotator is by far the most effective control procedure (compare Fig.~\ref{fig11} with
Fig.~\ref{fig10}(a)). Strikingly, solely decreasing the impulse of the pulses acting on
the hub $\left(  M=0\right)  $ is a much better choice than controlling even
several peripheral rotators but not the hub, as in the case $M=5=N/2$ shown in
Fig.~\ref{fig10}(b). The reason for this relatively good effectiveness stems from two
facts. First, solely controlling the hub does not significantly break the
synchronization of the whole network when $N$ is sufficiently large (as for
$N=10$, cf. Fig.~\ref{fig11}). Second, its maximum degree of connectivity allows the
hub to directly influence \textit{all} the remaining (peripheral)
rotators$-$in the sense of taming their chaotic dynamics$-$due to it is
behaving as an energy sink for sufficiently narrow pulses, as seen in the
energy analysis above (cf. Sec.~II~B).

\subsection{Control on both the central and the peripheral rotators}

\begin{figure}[tbp]
\begin{tabular}{cc}
\includegraphics[width=4cm]{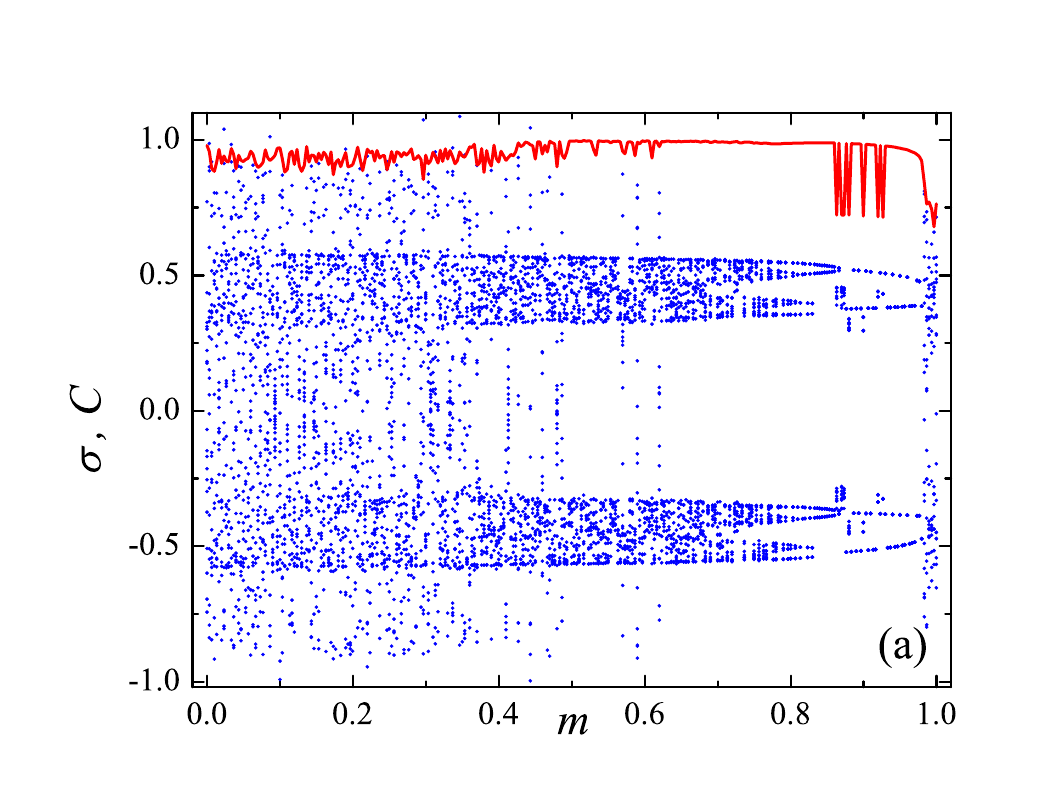} &
\includegraphics[width=4cm]{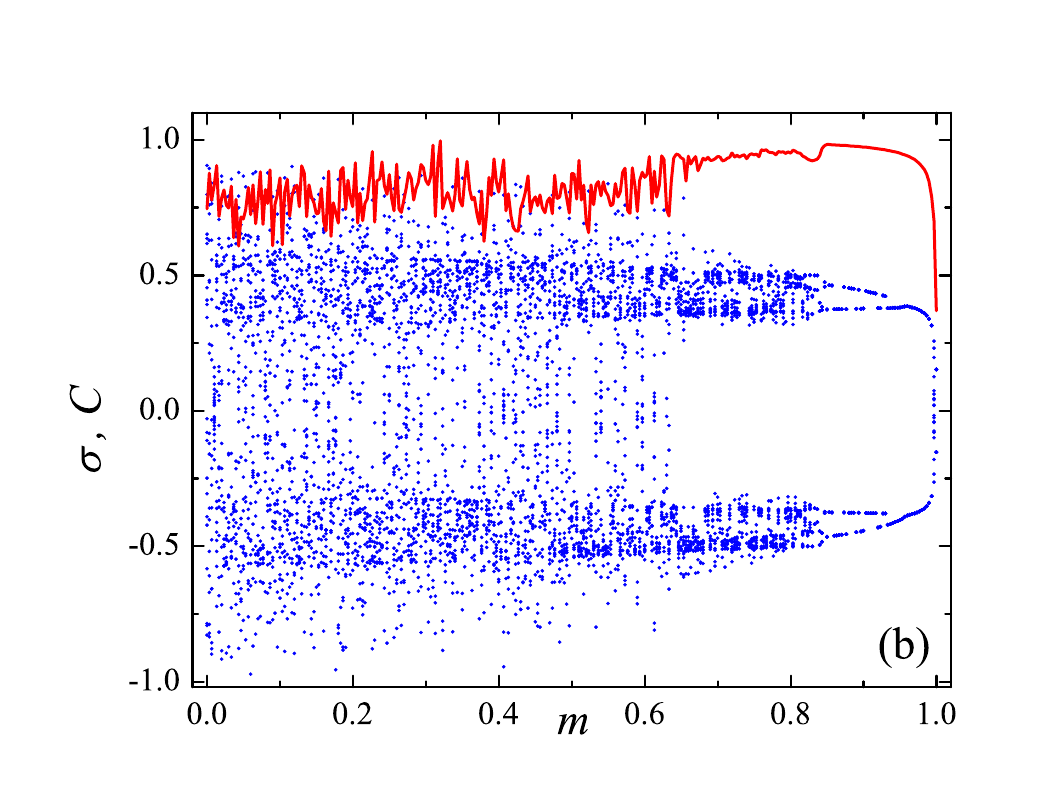} \\
\includegraphics[width=4cm]{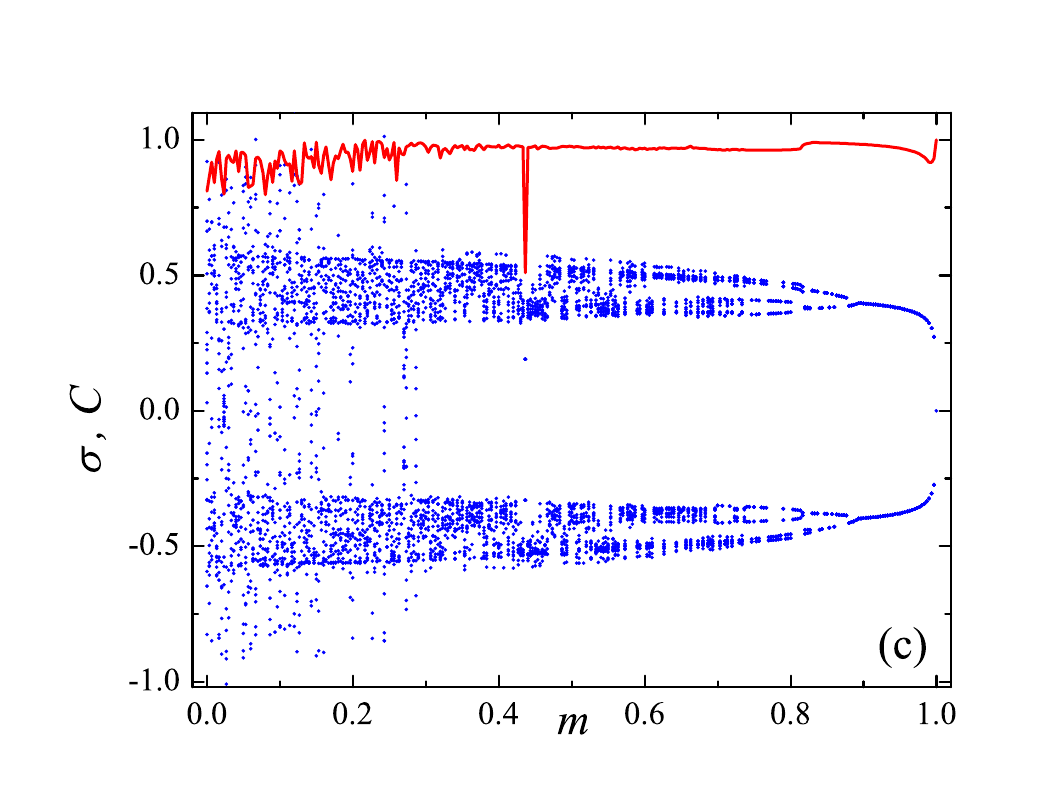} &
\includegraphics[width=4cm]{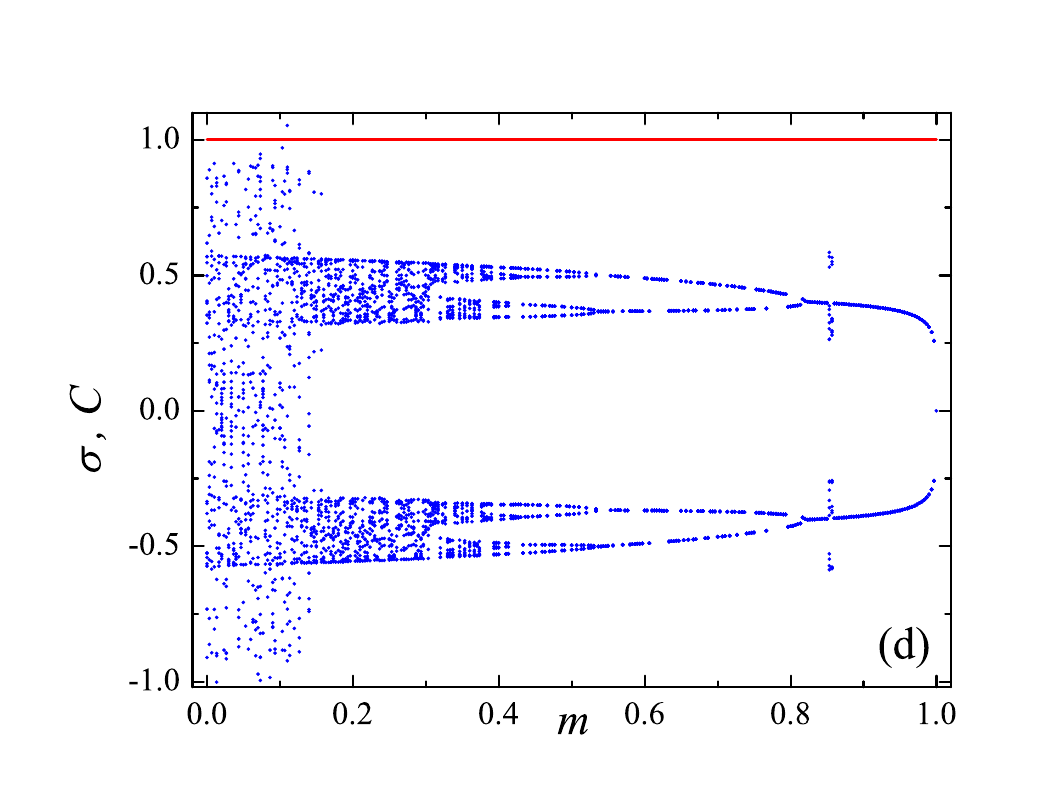} \\
\end{tabular}%
\caption{(Color online) Bifurcation diagrams of the average velocity $\sigma$
(blue (black) dots) and correlation function $C$ (red (gray) line) as a
function of the shape parameter $m=m_{j}=m_{H}$ for the case of a central
rotator subjected to impulse control, $N=10,\lambda=0.1,\delta=0.2,T=5.52$,
and four values of the number of peripheral rotators $x_{j}$ subjected to
impulse control: (a) $M=1$, (b) $M=5$, (c) $M=7$, and (d) $M=9$. The
quantities plotted are dimensionless.}
\label{fig12}
\end{figure}

Once the hub has been subjected to impulse control, one could expect \textit{a
priori} that additionally controlling other (peripheral) rotators should
improve the network's regularization. When the impulse transmitted by the
control pulses is comparable to that transmitted by the trigonometric pulses
(see Fig.~\ref{fig4}), one typically finds the opposite effect however: a deterioration
of the network's regularization, as for the case $M=1$ (cf. Fig.~\ref{fig12}(a)). This
deterioration effect, which occurs together with increasing desynchronization,
persists, and even increases, as the number of peripheral rotators subjected
to control is increased, as for the case $M=5=N/2$ (cf. Fig.~\ref{fig12}(b)) where
desynchronization is maximum (compare Fig.~12(b) with Figs. \ref{fig12}(a) and \ref{fig12}(c)
which correspond to the cases $M=1$ and $M=7$, respectively). As expected,
when all rotators are subjected to the same impulse control, as for the case
$M=9$ (cf. Fig.~\ref{fig12}(d)), the network synchronization becomes perfect and the
regularization route as the shape parameter is varied coincides with that of
an isolated rotator subjected to the same remaining parameters, involving
several consecutive crises followed by an inverse period doubling to finally
reach the equilibrium $\left(  x=0,\overset{.}{x}=0\right)  $ when the common
shape parameter is sufficiently near $1$.

\section{DISCUSSION}

To summarize, we have demonstrated theoretically and numerically that the
impulse transmitted by periodic pulses is a fundamental quantity for the
reliable control of the chaotic behavior of starlike networks of damped kicked
rotators. We have shown how the effectiveness of pulse reshaping, when it is
applied to a single node, strongly depends upon the degree of the target node:
applying impulse-decreasing pulses to the highest-degree node is by far the
best suppressory strategy, while applying them to low-degree nodes is the
poorest choice. In the case of applying pulse control to several nodes, we
found rather counterintuitive results: more is not only different but often
means poorer regularization. We have shown that this is due to
desynchronization phenomena which result from the competition between two
comparable populations of rotators subjected to pulses transmitting comparable
but different impulses. When the pulses' impulse is sufficiently small, the
rotator subjected to control behaves as an energy sink for the remaining
rotators, and this is ultimately the basic physical mechanism leading to the
network's regularization. Clearly, the effectiveness of this localized
dissipation of energy strongly depends upon the target node's degree, which
explains why controlling the central node is a much better choice than
controlling a peripheral node. The decreasing-impulse-induced chaos-control
scenario discussed could find applications in diverse biological coupled
systems [31], including neuronal networks [32]. It may also be useful to
optimally control chaos in scale-free networks of dissipative
periodically-kicked oscillators since a highly connected node in such a
network can be thought of as a hub of a locally starlike part of the network,
with a degree of connectivity that belongs to the complete network's
degree-of-connectivity distribution.

\begin{acknowledgments}
Useful discussions with Pedro J. Mart\'{\i}nez and \'{A}ngel Mart\'{\i}nez
Garc\'{\i}a-Hoz are gratefully acknowledged. R.C. gratefully acknowledges
financial support from the Ministerio de Econom\'{\i}a y Competitividad
(MINECO, Spain) through Project No. FIS2012-34902 cofinanced by FEDER funds,
and from the Junta de Extremadura (JEx, Spain) through Project No. GR15146.
\end{acknowledgments}


\begin{thebibliography}{99}                                                                                               %


\bibitem {1}Y.-Y. Liu, J.-J. Slotine, and A.-L. Barab\'{a}si, Nature
\textbf{473}, 167 (2011).

\bibitem {2}T. Nepusz and T. Vicsek, Nat. Phys. \textbf{8}, 568 (2012).

\bibitem {3}M. P\'{o}sfai, Y.-Y. Liu, J.-J. Slotine, and A.-L. Barab\'{a}si,
Sci. Rep. \textbf{3}, 1067 (2013).

\bibitem {4}S. P. Cornelius, W. L. Kath, and A. E. Motter, Nat. Commun.
\textbf{4}, 1942 (2013).

\bibitem {5}G. Menichetti, L. Dall'Asta, and G. Bianconi, Phys. Rev. Lett.
\textbf{113}, 078701 (2014).

\bibitem {6}R. Laje and D. V. Buonomano, Nat. Neurosci. \textbf{16}, 925 (2013).

\bibitem {7}D. Delpini \textit{et al}., Sci. Rep. \textbf{3}, 1626 (2012).

\bibitem {8}G. Chen and X. Dong, \textit{From Chaos to Order} (World
Scientific, Singapore, 1998).

\bibitem {9}R. Chac\'{o}n, \textit{Control of Homoclinic Chaos by Weak
Periodic Perturbations} (World Scientific, Singapore, 2005).

\bibitem {10}\textit{Handbook of Chaos Control}, 2nd ed., edited by E.
Sch\"{o}ll and H. G. Schuster (Wiley-VCH, Weinheim, 2008).

\bibitem {11}W. Wang, I. Z. Kiss, and J. L. Hudson, Phys. Rev. Lett.
\textbf{86}, 4954 (2001).

\bibitem {12}P. J. Mart\'{\i}nez and R. Chac\'{o}n, Phys. Rev. Lett.
\textbf{93}, 237006 (2004); \textbf{96}, 059903(E) (2006).

\bibitem {13}K. Rajan, L. F. Abott, and H. Sompolinsky, Phys. Rev. E
\textbf{82}, 011903 (2010).

\bibitem {14}R. Albert and A.-L. Barab\'{a}si, Rev. Mod. Phys. \textbf{74}, 47 (2002).

\bibitem {15}L. M. Pecora, Phys. Rev. E \textbf{58}, 347 (1998).

\bibitem {16}Z. Ma, G. Zhang, Y. Wang, and Z. Liu, J. Phys. A: Math. Theor.
\textbf{41}, 155101 (2008).

\bibitem {17}A. Bergner \textit{et al}., Phys. Rev. E \textbf{85}, 026208 (2012).

\bibitem {18}P. V. Kuptsov and A. V. Kuptsova, Phys. Rev. E \textbf{92},
042912 (2015).

\bibitem {19}D. A. Steck, W. H. Oskay, and M. G. Raizen, Science \textbf{293},
274 (2001).

\bibitem {20}R. Chac\'{o}n, Phys. Rev. E \textbf{74}, 046202 (2006).

\bibitem {21}R. Chac\'{o}n and A. Mart\'{\i}nez Garc\'{\i}a-Hoz, Phys. Lett. A
\textbf{281}, 231 (2001).

\bibitem {22}R. Chac\'{o}n and A. Mart\'{\i}nez Garc\'{\i}a-Hoz, Phys. Rev. E
\textbf{68}, 066217 (2003).

\bibitem {23}A. Koseska, E. Volkov, and J. Kurths, Phys. Rep. \textbf{531},
173 (2013).

\bibitem {24}V. K. Melnikov, Trans. Moscow Math. Soc. \textbf{12}, 1 (1963)
[Tr. Mosk. Ova. \textbf{12}, 3 (1963)].

\bibitem {25}J. Guckenheimer and P. Holmes, \textit{Nonlinear Oscillations,
Dynamical Systems, and Bifurcations of Vector Fields} (Springer-Verlag, New
York, 1983).

\bibitem {26}G. Benettin, L. Galgani, and J. M. Strelcyn, Phys. Rev. A
\textbf{14}, 2338 (1976); I. Shimada and T. Nagasama, Prog. Theor. Phys.
\textbf{61}, 1605 (1979).

\bibitem {27}I. S. Gradshteyn and I. M. Ryzhik, \textit{Table of Integrals,
Series, and Products} (Academic Press, San Diego, 1980).

\bibitem {28}J. V. Armitage and W. F. Eberlein, \textit{Elliptic Functions}
(Cambridge University Press, Cambridge, 2006).

\bibitem {29}See, e.g., S. H. Strogatz, \textit{Nonlinear Dynamics and Chaos}
(Addison-Wesley, Reading, MA, 1994), p. 249.

\bibitem {30}A. N. Pisarchik and U. Feudel, Phys. Rep. \textbf{540}, 167 (2014).

\bibitem {31}E. Ullner, A. Zaikin, E. I. Volkov, and J. Garc\'{\i}a-Ojalvo,
Phys. Rev. Lett. \textbf{99}, 148103 (2007).

\bibitem {32}S. J. Schiff, \textit{Neural Control Engineering} (MIT Press,
Cambridge, 2012).
\end{thebibliography}
\end{document}